\documentclass[manuscript]{emulateapj}

\def\bc{\begin{center}}
\def\ec{\end{center}}                                                                                            

\shorttitle{Structural Properties of UCDs}
\shortauthors{Evstigneeva et al.}

\begin{document}


\title{Structural Properties of Ultra-Compact Dwarf Galaxies \\ in the Fornax and Virgo Clusters}


\author{E.~A.~Evstigneeva, M.~J.~Drinkwater}
\affil{Department of Physics, University of Queensland, QLD 4072, Australia; e.evstigneeva@gmail.com, m.drinkwater@uq.edu.au}
\author{C.~Y.~Peng}
\affil{NRC Herzberg Institute of Astrophysics, 5071 West Saanich Road, Victoria, BC, Canada V9E 2E7; cyp@nrc-cnrc.gc.ca}
\author{M.~Hilker}
\affil{European Southern Observatory, Karl-Schwarzschild-Str. 2, 85748 Garching bei M\"unchen, Germany; mhilker@eso.org}
\author{R.~De~Propris}
\affil{Cerro Tololo Inter-American Observatory, Casilla 603, La Serena, Chile; rdepropris@ctio.noao.edu}
\author{J.~B.~Jones}
\affil{Astronomy Unit, School of Mathematical Sciences, Queen Mary University of London, Mile End Road, London E1 4NS; bryn.jones@qmul.ac.uk}
\author{S.~Phillipps}
\affil{Astrophysics Group, Department of Physics, University of Bristol, Tyndall Avenue, Bristol BS8 1TL; S.Phillipps@bristol.ac.uk}
\and
\author{M.~D.~Gregg\altaffilmark{1}, A.~M.~Karick\altaffilmark{1}}
\affil{Department of Physics, University of California, Davis, CA 95616, USA; gregg,akarick@igpp.ucllnl.org}


\altaffiltext{1}{Institute for Geophysics and Planetary Physics, Lawrence Livermore National Laboratory, L-413, Livermore, CA 94550, USA}


\begin{abstract}
We present a detailed analysis of high-resolution two-band {\em Hubble Space Telescope} 
Advanced Camera for Surveys imaging of 21 ultra-compact dwarf (UCD) galaxies in 
the Virgo and Fornax Clusters. 
The aim of this work is to test two formation hypotheses for
UCDs---whether they are bright globular clusters (GCs) or stripped (``threshed'') early-type dwarf
galaxies---by direct comparison of UCD structural parameters and
colors with GCs and galaxy nuclei.
We find that the UCD surface brightness profiles can be described by a
range of models and that the luminous UCDs in particular can not be
described by standard King models with tidal cutoffs as they have extended outer halos.
This is not expected from traditional King models of GCs, 
but is consistent with recent results for massive GCs.
The total luminosities, colors and sizes of the UCDs (their position in the 
color-magnitude and luminosity-size diagrams) 
are consistent with them being either luminous GCs or threshed nuclei of both early-type   
and late-type galaxies (not just early-type dwarfs). 
For the most luminous UCDs we estimate color gradients over a limited range 
of radius. These are systematically positive in the sense of getting
redder outwards: mean $\Delta(F606W-F814W) = 0.14$ mag per 100 pc with
$rms = 0.06$ mag per 100 pc. 
The positive gradients found in the bright UCDs are consistent with them being
either bright GCs or threshed early-type dwarf galaxies (except VUCD3). 
In contrast to the above results we find a very significant
($>99.9\%$ significance) difference in the sizes of UCDs and early-type 
galaxy nuclei: the effective radii of UCDs are $2.2_{-0.1}^{+0.2}$ times larger than
those of early-type galaxy nuclei at the same luminosity. This result
suggests an important test can be made of the threshing hypothesis by
simulating the process and predicting what size increase is expected.
\end{abstract}





\section{Introduction}

Ultra-compact dwarf (UCD) galaxies are a class of stellar system
originally discovered in the Fornax Cluster (Hilker et al.\ 1999,
Drinkwater et al.\ 2000).  They appear star-like in ground-based
photographic survey images, but have recession velocities consistent
with cluster membership. UCDs have spectra typical of old stellar
populations but they are generally far more luminous than ordinary Milky
Way (MW) globular clusters (GCs), and much more compact than similarly
luminous dwarf spheroidal galaxies.

UCDs have now been detected in the Virgo (Ha\c{s}egan et al.\ 2005,
Jones et al.\ 2006) and Centaurus Clusters (Mieske et al.\ 2007) and,
possibly, in Hydra\,I (Wehner \& Harris 2007) and Abell\,1689 (Mieske
et al.\ 2004). On the other hand, they are much less common in groups
(e.g. Evstigneeva et al.\ 2007a) and the general field (Liske et al.\
2006), suggesting that UCDs are somehow related to the cluster
environment. We have now surveyed the central region of the Fornax
Cluster for less luminous objects: we find that the central region of
the cluster contains a large population of UCDs (60 objects to a magnitude
limit of $b_J=21.5$ or roughly $M_V\sim-10.3$; Drinkwater et al.\ 2004, Gregg et al.\
2008) although at such faint limits they clearly overlap the GC
populations associated with the central cluster galaxies.

There are three main scenarios for the origin of UCDs. (1) They may
simply be luminous GCs, encountered near giant
elliptical galaxies because such systems often possess populous
GC systems (Mieske et al.\ 2002). (2) A variant of this is
that the bright GCs, and UCDs, may be formed via the
merger of super star clusters which are abundant in nearby galaxy
mergers (Fellhauer \& Kroupa 2002). Presumably the merger process is very
important in the formation of bright elliptical galaxies in
clusters. (3) Finally, they may be the stripped (``threshed'') nuclei of former
nucleated early-type dwarf galaxies, whose envelopes have been tidally removed 
(Bekki et al.\ 2001). 

Part of the motivation for the stripping (``threshing'') hypothesis is
the similarity between UCDs and the nuclei of dwarf elliptical
galaxies. Our initial study of the internal dynamical properties
of UCDs (Drinkwater et al.\ 2003)
established that the UCDs are more closely related to dwarf galaxy
nuclei than to GCs.  More recently we have been able to include more
luminous extra-galactic GCs in the analysis, suggesting that there is
a continuum of dynamical properties from GCs to UCDs (Evstigneeva et
al.\ 2007, Hilker et al.\ 2007).  While this supports the
interpretation that UCDs are simply bright GCs, it does not suffice to
distinguish between the three above scenarios. It is only by a
thorough comparison of structure, dynamics and stellar populations
(e.g. Evstigneeva et al.\ 2007) between UCDs and their supposed
progenitors that we can separate the main pathway of UCD formation.
We need to bear in mind, however, that there may be more than one
channel involved.

In this present paper we focus specifically on the structural
parameters of UCDs which can be compared to possible progenitors. The
work we present here was motivated by the large sample of fainter UCDs
we identified in the Fornax Cluster (Drinkwater et al.\ 2004, Gregg et
al.\ 2007). This sample was ideally suited to a {\em Hubble Space
  Telescope} (HST) ``snapshot'' program, allowing us to obtain images
of a statistically useful subset of our entire sample. Our primary aim
was to test the first and the third hypotheses above by direct
comparison of UCD structural parameters with GCs and galaxy nuclei
also measured by HST.

The structure of the paper is as follows. In Section~2
we describe the data sample and in Section~3 we describe
the HST imaging and image modeling. In Section~4 we present analysis of the
structure and colors of the most extensive and complete sample of UCDs
in the Fornax and Virgo Clusters observed to date. The sample includes
six Virgo UCDs initially presented by Evstigneeva et al.\ (2007), as
we have made several improvements to the image analysis. In
Section~5 we summarize our results and findings. Throughout this paper we
adopt distance moduli of 30.92 for the Virgo Cluster and 31.39 for the
Fornax Cluster (Freedman et al.\ 2001), corresponding to the distances of 
15.28 Mpc to Virgo and 18.97 Mpc to Fornax.

\section{UCD galaxy sample}

In this section we briefly describe the properties we use to define
UCD galaxies and how we obtained the sample for our HST
observations. More details are given in the respective discovery
papers listed below.

\subsection{Definition of the UCD galaxy type}

As noted above, the UCD galaxies originally discovered in the Fornax
Cluster were unresolved in ground-based photographic imaging, but had
redshifts consistent with cluster membership. The first UCDs found
were much more luminous than any known GCs, with
apparent magnitudes in the range $17.7<b_J<19.7$ (Drinkwater et al.\
2000) or $-13.4<M_V<-11.9$. At these luminosities the UCDs were clearly distinct from any
known galactic or stellar system (Drinkwater et al.\ 2003). Note that,
although the ``unresolved'' criterion depends on the image quality,
there is a very large gap in parameter space between the UCDs and
normal dwarf galaxies, so it only serves to remove clearly normal
galaxies from the samples. We have
since extended our searches to fainter limits ($b_J<21.5$  or roughly $M_V<-10.3$ in the Fornax
Cluster; Drinkwater et al.\ 2004, Gregg et al.\ 2008): at these limits there is a clear overlap
with objects that would normally by classified as GCs.

For the purposes of the current discussion we will use the term
``UCD'' to include all the compact (i.e.\ unresolved in our
ground-based photographic imaging) intra-cluster objects we have
discovered in the Virgo and Fornax Clusters, recognizing that there is
clear overlap at the fainter limits with objects conventionally
classified as GCs.

\subsection{UCD targets}

As noted above, our HST snapshot proposal was motivated by the
availability of a large sample of UCDs with a range of luminosities
from the Fornax and Virgo Clusters. Although the UCDs were
originally discovered (Drinkwater et al.\ 2000) through the
``all-object'' approach of the Fornax Cluster Spectroscopic Survey
(Drinkwater et al.\ 2000a), our subsequent UCD searches were more
selective. In particular we imposed color selection to avoid the reddest
stellar objects as none of the original UCDs were this red.

In the Virgo Cluster we carried out a targeted search specifically
aimed at detecting {\em luminous} UCDs with similar properties to the originally discovered 
Fornax UCDs. This was very successful with 9 UCDs detected in just a few
hours of observing time (Jones et al.\ 2006). The Virgo objects were
selected in the magnitude range $16.0<b_J<20.2$ (roughly $-15.3<M_V<-11.1$) and the color range
$b_J-r_F<1.6$ (roughly $V-I<1.5$). Our spectroscopic observations in Virgo were about 65\% 
complete, so we estimate the true population of bright UCDs
in the central 1-degree ($\sim 270$ kpc) radius region of the Virgo Cluster to be
about 14.

In the Fornax Cluster we extended the search to much fainter limits using very similar approach. 
In our final observations with the 2dF system we selected unresolved objects in the
magnitude range $16.0<b_J<21.5$ (roughly $-15.8<M_V<-10.3$) and the color range $b_J-r_F<1.7$ (roughly $V-I<1.6$) and
we limited the area searched to a 0.9-degree radius from the center of
the Fornax Cluster. This approach was again very successful with a
total of 60 UCDs detected in these limits (Gregg et al.\ 2008).
Allowing for the incompleteness of our spectroscopic
observations, we estimate the true population of UCDs in the central
0.9-degree ($\sim 300$ kpc) region of the Fornax Cluster to be about 105.
Note that this is to a fainter limit ($b_J<21.5$) than our Virgo sample 
($b_J<20.2$).

In both cases we only selected targets that were classified as
unresolved or merged with another object from our ground-based
photographic survey imaging (see Drinkwater et al.\ 2000a). At these
magnitudes most ``merged'' objects consisted of a stellar object with
one or more faint companions. This selection would not remove any UCDs
from the samples.

We selected 50 of the known Fornax and Virgo UCDs to observe with HST. The targets were
chosen to cover a range of luminosities, as well as avoiding
overlap with our previous observations of the first Fornax UCDs
(program 8685). One object from this earlier program was reselected:
Fornax UCD3. We chose to reobserve this object because of its complex
morphology we hoped to better resolve with the new observations.

\section{HST observations and Image modeling}

We obtained images of 21 of the requested Fornax and Virgo UCDs in the course 
of HST snapshot program 10137.  The data were taken with the Advanced Camera for
Surveys (ACS), High Resolution Channel (HRC), through the F606W and
F814W filters. Exposure times were 870 sec in F606W and 1050 sec in
F814W.  The HRC scale is $0.025''$ pixel$^{-1}$. For the image analysis
we used {\sc MultiDrizzle}\footnote{See http://stsdas.stsci.edu/multidrizzle.}
(*.mdz) files retrieved from the HST archive.

To measure the total magnitudes, we plotted curves of growth (integrated magnitude 
vs.\ circular aperture radius)
to find an aperture radius large enough to enclose all the light from an object.
The instrumental F606W and F814W magnitudes were transformed into
Landolt $V$ and $I$ band following Sirianni et al.\ (2005).  The
resulting $V$ magnitudes and $V - I$ colors are listed in Table~1.

The images of Fornax and Virgo UCDs were modeled using the two-dimensional
fitting algorithm {\sc GALFIT} (Peng et al.\ 2002) and assuming empirical
King, S\'{e}rsic and Nuker models for the luminosity profile. 

The empirical King profile is characterized by the core radius, $R_c$, and the
tidal radius, $R_t$, and has the following form (Elson 1999):
\begin{equation}
I(R) = I_0 \left[\frac{1}{(1+(R/R_c)^2)^{\frac{1}{\alpha}}} - \frac{1}{(1+(R_t/R_c)^2)^{\frac{1}{\alpha}}} \right]^{\alpha},
\end{equation}
where I$_0$ is the central surface brightness. We tried both the
standard model with $\alpha = 2$ and generalized model with variable
$\alpha$.  

The S\'{e}rsic power law has the following form (S\'{e}rsic 1968):
\begin{equation}
I(R) = I_{\rm eff} \,\, exp \left[-k \left(\left(\frac{R}{R_{\rm eff}}\right)^{\frac{1}{n}} - 1 \right) \right],
\end{equation}
where $R_{\rm eff}$ is the half-light (effective) radius, $I_{\rm
  eff}$ is surface brightness at the effective radius, $n$ is the
concentration parameter ($n=4$ for de Vaucouleurs profile and $n=1$ for
exponential profile) and $k$ is a constant which depends on $n$.

The Nuker law is as follows (Lauer et al. 1995):
\begin{equation}
I(R) = I_b \,\, 2^{\frac{\beta - \gamma}{\alpha}} \, \left(\frac{R}{R_b} \right)^{- \gamma} \, 
\left[1 + \left( \frac{R}{R_b} \right)^{\alpha} \right]^{\frac{\gamma - \beta}{\alpha}}. 
\end{equation}
It is a double power law, where $\beta$ is the outer power law slope,
$\gamma$ is the inner slope, and $\alpha$ controls the sharpness of
the transition (at the ``break'' radius $R_b$) from the inner to the
outer region. $I_b = I(R_b)$. 

The UCDs are barely resolved---even with the HST/ACS resolution---so
to obtain their intrinsic luminosity profiles,
we must correct for the telescope point-spread function (PSF).  
We derived artificial PSFs for the images in each filter using the {\sc TinyTim}
software\footnote{See http://www.stsci.edu/software/tinytim.} and {\sc
  {\sc MultiDrizzle}} as described in Evstigneeva et al.\ (2007).  
The size of the PSFs was chosen to be $3.5''\times3.5''$ ($140 \times 140$ pixel), 
slightly larger than the minimum size recommended by {\sc TinyTim}. 
For the brightest UCDs, which are also the most extended, we used PSFs of a larger 
size (the same as for the color gradient analysis in Section~4.3.1) to correct for the extended 
PSF halo in the F814W filter. 

{\sc GALFIT} models an analytic profile convolved with the PSF and determines the
best-fitting profile parameters by minimizing residuals between the
model and original two-dimensional image.
We limited all the models to a maximum fitting radius defined by the point where the UCD light profile 
reaches the background noise level in the $V$ image (the models are not constrained beyond that point). 
The image of Fornax UCD3 is complicated by the presence of a background object in projection 
(possibly a spiral galaxy, see Evstigneeva et al.\ 2007).
We therefore restricted the model fitting to the half of the image
least affected by the background source.

The sky (background) was estimated and subtracted from the UCD images before running {\sc GALFIT}. 
The sky was initially subtracted  by {\sc MultiDrizzle}. 
We then applied additional background corrections, determined from empty parts of the images.
So we held the sky value fixed to zero when fitting the images. 
It is important to hold the sky fixed when fitting standard models to an object, because 
the model function used may not be optimal and the model mismatch can push the sky around a little.
On the other hand, if the object is fitted as well as possible (by multi-component models) and 
if the sky region is large enough to fit, then the sky can be allowed to vary as a free parameter.  
This is what we did for the color gradient analysis in Section~4.3.1.
Thus, in this section, for the structural modeling, we held the sky value fixed (did not fit the sky with 
{\sc GALFIT}). We, however, did the tests of changing the sky by hand (subtracting the sky values found 
by GALFIT in Section~4.3.1) and redoing the fit. It did not affect the structure parameters 
in Tables~1 \& 2 very much: the changes were within the uncertainties given in the tables. 
However, the sky corrections can be critical for the outer color profile, for the tiny color 
gradients we find in UCDs.

The quality of the {\sc GALFIT} model fits is shown in Figure~1. For this figure 
we used the {\sc ELLIPSE} task in {\sc IRAF} to produce 
one-dimensional surface brightness profiles for the objects and
(PSF-convolved) {\sc GALFIT} models.

To choose the {\it best model} for each object (see the last column of Table~2), 
we used $\chi_{\nu}^2$ values of the fits (Peng et al.\ 2002). 
In the case of faint UCDs ($m_{V,0} \sim 19.9 - 20.9$ mag), there is no preference of one model over another: 
all the models (Nuker, King and S\'{e}rsic) seem to fit the UCDs equally well within the errors.  
The bright UCDs ($m_{V,0} \sim 17.5 - 18.9$ mag) have extended outer halos and appear to be best fitted by 
a double power law (Nuker) or two-component models (the central component was fitted by King with $\alpha=2$ 
or S\'{e}rsic and the outer envelope by S\'{e}rsic). 
This does not necessarily mean that bright and faint UCDs are intrinsically different. 
If we had deeper observations for the fainter UCDs, 
we would possibly be able to detect outer halos in them were any present. 
The detection of extended halos around the luminous UCDs is not
consistent with their having the standard King profiles traditionally
associated with GCs, but recent work of McLaughlin \&
van der Marel (2005) and McLaughlin et al.\ (2008) has shown that extended halos are a general
characteristic of massive GCs in the Milky Way and some of its satellites and NGC5128.

In Table~1 we quote effective radius $R_{\rm eff}$, model magnitude $M_{V,0}^{\rm mod}$ and 
ellipticity $\epsilon$. 
The ellipticity value is the best model value (see last column of Table~2).
The $R_{\rm eff}$ and $M_{V,0}^{\rm mod}$ values were obtained from generalized King 
(or standard King with $\alpha=2$, if it fits better) 
models for one-component UCDs and King+S\'{e}rsic models for two-component UCDs, 
via numerical integration (to infinity) of the $V$ and $I$ luminosity profiles. 
These models give the most stable estimates for $R_{\rm eff}$  
as discussed in Evstigneeva et al.\ (2007). 
The $M_{V,0}^{\rm mod}$ values are only slightly different from the observational $M_{V,0}$ values, 
obtained by integrating the actual image pixel values.    
In further analyses we use observational magnitudes ($M_{V,0}$). 
The choice of magnitude, however, does not change our final results and conclusions   
(e.g. equations (4)--(7) stay the same, as well as the size difference between UCDs and nuclei, found below).  
Table~1 also contains the parameters for the four bright Fornax UCDs from Evstigneeva et al.\ (2007) (HST/STIS data), 
as we have made some improvements to the image analysis and we use these data in the analyses below. 

In Table~2 we list more model parameters such as S\'{e}rsic index $n$;
King central surface brightness $\mu_{V,0}$, core radius $R_c$,
concentration $c$ and $\alpha$ parameter; Nuker inner slope $\gamma$,
outer slope $\beta$ and break radius $R_b$.  We do not list all the
parameters for all the models, but only the most significant ones
needed for qualitative analysis, mainly to compare UCDs with each
other. We recommend caution in using the actual values of these parameters 
in detailed analyses. One of the main reasons for this is that the objects are barely resolved. 
For example, for about half of the UCDs, the King
model fits appear good, but the core radii of these models $R_c <
1$ pixel. It means that these UCDs either have cores, which are unresolved, 
or do not have cores at all, so that the actual $R_c$ values are
uncertain. As a result, we can not trust the values of central surface
brightness and concentration obtained from these King models.  
The central surface brightness of UCDs can also be derived from S\'{e}rsic models: we 
can calculate $\mu_{V,0}$ from {\sc GALFIT}'s $m_{V,tot}$, $R_{\rm eff}$ and $n$ values  
analytically by using formulae for the S\'{e}rsic function.    
However, for models with high S\'{e}rsic indices ($n>2$), the calculations result in
unrealistically high central surface brightness values.  As for the Nuker parameters,
they are not always stable. If we change the radial extent to which we
fit this model, there is a good chance that all the parameters will
change considerably (Graham et al.\ 2003).

\section{Analysis and Discussion}

\subsection{Structural parameters}

The results of the modeling show that the UCDs have a range of 
S\'{e}rsic indices $n$ and King concentrations $c$, as well as 
a range of central slopes (seen from Nuker inner slopes $\gamma$): 
from flat ``King'' cores to central cusps. This suggests that 
convolving some cuspy models with the PSF allows such models 
to fit the seeing-blurred centers of some UCD profiles. 
We, however, can not make very strong conclusions regarding central 
cusps or cores in the UCDs, taking into account the limits of our data 
(described in the previous section).  

We tried to look for correlations of the model parameters in Table~2 
with the UCD luminosity, size and color, but did not find any.

We compared the distribution of UCD ellipticities (in Table~1) with those for MW GCs (Harris 1996), 
NGC5128 GCs (Holland et al.\ 1999, Harris et al.\ 2002) and M31 GCs (Barmby et al.\ 2007).
The two-sample K-S test shows that the UCD ellipticities are consistent with extragalactic GC 
distributions (NGC5128 and M31 GCs), but significantly different from 
the MW GC distribution. The Wilcoxon test gives the same result. 
It is interesting to note that Harris et al.\ (2002) found the MW GC ellipticity distribution 
to be significantly different from the NGC5128 and M31 GC distributions. 
Harris et al.\ (2002), however, do not place too much weight on their result because of 
uncertainties about possible selection effects and different methods for ellipticity measurements      
(see also Barmby et al.\ 2007).  
   
No correlation of ellipticity with luminosity, size or color was found for UCDs. 

In Figure~2 we present the luminosity-size diagram for UCDs. For comparison, we 
also show ``dwarf globular transition 
objects'' from Ha\c{s}egan et al.\ (2005), early-type galaxy nuclei 
from C\^ot\'e et al.\ (2006) and GCs (see caption of Figure~2).
The UCDs and GCs form a continuous distribution across the plane, but  
UCDs seem to be different to {\em typical} GCs in the sense that the
UCDs have sizes correlated with luminosities whereas the GCs do not. 
This difference has been reported previously (e.g. by Ha\c{s}egan et al.\ 2005 for 
``dwarf globular transition objects'' and five bright Fornax UCDs).

We obtain the following luminosity-size relation for the UCDs in our sample (fitting a linear least squares regression  
to the data for both Fornax and Virgo UCDs in Table~1): 
\begin{eqnarray}
{\rm log}\,R_{\rm eff} & = & -3.03 (\pm 0.55) - 0.35 (\pm 0.05) \, M_V \\
{\rm or} \,\,\, R_{\rm eff} & \propto & L_V^{0.88 \pm 0.13} \, .
\end{eqnarray}
If we exclude the two brightest objects, which may be different from all other UCDs (they are much brighter than 
all other UCDs and have the largest envelopes), we obtain:
\begin{eqnarray}
{\rm log}\,R_{\rm eff} & = & -2.11 (\pm 0.61) - 0.27 (\pm 0.05) \, M_V \\
{\rm or} \,\,\, R_{\rm eff} & \propto & L_V^{0.68 \pm 0.13} \, .
\end{eqnarray}

The fact that the UCDs show a luminosity-size relation while the GCs do not 
is not a result of any selection effects. 
To illustrate this point, we show in Figure~2 the selection boundaries for 
our UCD sample. The sample selection is only restricted at the faint magnitude end 
(by our survey flux limit corresponding to $M_V=-10.5\,{\rm mag}$, vertical line). 
The size limit (horizontal lines) is an approximate representation
of the minimum size that could be parametrized from HST/ACS imaging:
below this limit objects become unresolved. It does not represent a detection limit,
because the 2dF surveys were sensitive to objects regardless of their sizes.   
All the UCDs, however, are resolved with HST/ACS ($R_{\rm eff} > 1$ pixel).  

The luminosity-size relation we observe for the UCDs is similar to the {\em bright} GC
observations by Barmby et al.\ (2007). They report an increasing lower
bound on $R_{\rm eff}$ in the mass vs.\ $R_{\rm eff}$ plane for the
most massive GCs (masses $\geq 1.5 \times 10^6 \, {\rm M_{\odot}}$).
Barmby et al.\ interpreted it as an extension of a similar relation
for early-type galaxies following from the existence of a ``zone of
exclusion'' (ZOE) in the fundamental plane ($\kappa$-space), discussed
by Burstein et al.\ (1997).  According to Burstein et al.\ (1997), no
stellar system violates the rule $\kappa_1 + \kappa_2 < 8$, 
which means that the maximum global luminosity density of
stellar systems varies as mass$^{-4/3}$.  Assuming constant
mass-to-light ratios, this is equivalent to $R_{\rm min} \propto
L^{0.78}$, the same relation (within uncertainties) as we find for
UCDs.  So if we consider UCDs as a part of GC family, perhaps it would
be more correct to talk about the increasing lower boundary on $R_{\rm eff}$ 
rather than about luminosity-size relation for them, which
could be explained by the ZOE.  However, the existence of a ZOE for galaxies
is not quite understood yet.

The nuclei of early-type galaxies also show a luminosity-size
correlation as noted by C\^ot\'e et al.\ (2006).  Fitting a linear
least squares regression to all the nuclei brighter than
$M_V=-10.5\,{\rm mag}$, except for the brightest one at
$M_V\sim-16\,{\rm mag}$ (to approximately match the UCD luminosity
range), and excluding both unresolved and offset nuclei, we found the
following relation:
\begin{eqnarray}
{\rm log}\,R_{\rm eff} & = & -2.53 (\pm 0.58) - 0.28 (\pm 0.05) \, M_V \\
{\rm or} \,\,\, R_{\rm eff} & \propto & L_V^{0.70 \pm 0.13} \, , 
\end{eqnarray}
where $R_{\rm eff}$ is the mean of the two pass bands, $g'$ and $z'$.
This relation is very similar (especially the slope) to the UCD
luminosity-size relations above. This is consistent with the threshing
hypothesis for UCD formation from disrupted early-type galaxies.  We
do not show the nuclei of late-type galaxies (or ``nuclear star
clusters'') in Figure~2. They are nearly identical to the early-type
galaxy nuclei in the sense of luminosities and sizes (e.g.\ C\^ot\'e
et al.\ 2006 and references therein), so they would not add new
information to the figure. The similarity, however, 
means that UCDs---if formed by disruption---could be the remnant
nuclei of both early-type and late-type galaxies, not just dE,Ns as it was 
initially suggested.

From inspection of Figure~2 and the luminosity-size relations above we
note that the UCDs are significantly larger than the corresponding
early-type galaxy nuclei at the same luminosity. This is consistent
with an earlier comparison we made of the brighter Fornax UCDs with
some fainter nuclei (de Propris et al.\ 2005), but we are now able
to confirm the size difference is not due to the difference in
luminosity of those samples.

Taking the
luminosity-size relation fitted for the galaxy nuclei as a baseline,
we calculate that the UCDs are an average factor of 
$\Delta\log R_{\rm eff} = 0.35 \pm 0.03$ larger 
(we do not include the two brightest UCDs with the largest envelopes here). 
This difference is significant 
at a confidence level $>99.9\%$ (using the T-test) and corresponds to the
effective radii of the UCDs being on average $10^{0.35\pm0.03}=2.2_{-0.1}^{+0.2}$ 
times larger than the nuclei of the same luminosity. 
In Figure~3 we show the histograms of $R_{\rm eff}/L_V^{0.7}$ for nuclei and UCDs 
separately (in the same luminosity range as was used to derive relations (7) \& (9)), 
which better emphasizes the difference between UCDs and early-type galaxy nuclei. 
Although this observation is 
superficially inconsistent with a process whereby these galaxy
nuclei are stripped to form UCDs, the change in size may be a result
of the stripping process.  The nuclei may be more concentrated
(denser) than the UCDs due to a truncation effect caused by their
location within the prominent stellar envelope (or dark matter halo)
of a galaxy which does not affect the isolated UCDs.  Conversely, the
stripping process itself may well result in dynamical heating of the
nuclei, so they expand as their host galaxies are disrupted. The 
simulations of the stripping process by Bekki et al.\ (2001, 2003) do
indicate some expansion of the remnant core, but no quantitative
results are provided. Given our new observational result, it will be
important to test this in detail against simulations of the threshing
process to see if it is consistent with the size difference we have
found between UCDs and the early-type galaxy nuclei.
As an additional constraint on this process we note that the cores of
the largest UCDs (which have small stellar envelopes) have magnitudes
and sizes similar to the other UCDs (which do not have envelopes):
$-11.16 \geq M_V\geq-12.39\,{\rm mag}$, $6.5 \leq R_{\rm eff}\leq12.8\,{\rm pc}$.  
They are still larger on average than
early-type galaxy nuclei at the same luminosity.  This means that the
gravitational potential of these small stellar envelopes is not be
enough to squeeze the central component to the extent that the galaxy
nuclei are compressed.

\subsection{Color-magnitude diagrams}

The total colors of the UCDs from our ACS survey (Table~1) can be compared
to those of other hot stellar systems in the same environment. We therefore
constructed color-magnitude diagrams (CMDs) for GCs, nuclei of
early-type galaxies and UCDs in the Virgo and Fornax Clusters. These are
shown in Figures~4 \& 5. All magnitudes and colors were transformed to the 
Johnson-Cousins $V \& (V-I)$ system to facilitate the comparison of our data to 
many other works in this filter set.

The GC data were taken from the ACS Virgo Cluster Survey
(M87 and M49 GCs, C\^ot\'e et al.\ 2004, Peng et al.\ 2006) and the 
ACS Fornax Cluster Survey (NGC1399 and NGC1404 GCs, Jord\'an et al.\ 2007). 
The magnitudes and colors for the nuclei of Virgo early-type galaxies and 
``dwarf globular transition objects'' were also taken from the 
ACS Virgo Cluster Survey (C\^ot\'e et al.\ 2006, Ha\c{s}egan et al.\ 2005).
The transformations from the ACS survey $g'$ and $z'$ AB magnitudes to the
$V$ and $I$ magnitudes were performed by using theoretical single stellar population 
(SSP) models of Bruzual \& Charlot (2003). We derived transformation equations for 
three different age bins to account for possible intermediate age
populations in nuclei and ``dwarf globular transition objects''. For a Chabrier initial 
mass function (IMF) and a metallicity range of -2.2 to 0.6 dex, 
the transformation equations for the three age bins are as follows:
\begin{eqnarray}
{\rm 11-13 \,\, Gyr:} \,\,\, V_0 & = & g'_{AB}-0.004-0.301(g'-z')_{AB} \\
(V-I)_0 & = & 0.445+0.518(g'-z')_{AB} \, ;
\end{eqnarray}
\begin{eqnarray}
{\rm 6-8 \,\, Gyr:} \,\,\, V_0 & = & g'_{AB}-0.005-0.298(g'-z')_{AB}\\
(V-I)_0 & = & 0.449+0.515(g'-z')_{AB} \, ;
\end{eqnarray}
\begin{eqnarray}
{\rm 2-4 \,\, Gyr:} \,\,\, V_0 & = & g'_{AB}+0.015-0.315(g'-z')_{AB}\\
(V-I)_0 & = & 0.451+0.496(g'-z')_{AB} \, .
\end{eqnarray}
For the objects in Figures~4 \& 5, we chose the transformation equations 
obtained for the very old ages. The age effect on the transformations,
however, is not strong, as shown by the color-magnitude relations for the Virgo
nuclei in Figure~4, and the choice of age does not affect our main conclusions below.

The tick marks on the bottom of the CMDs in Figures~4 \& 5 indicate 
the peak colors for the blue and red GC populations  
in Virgo and Fornax derived by Larsen et al.\ (2001). 
They are in a good agreement with the average transformed colors for 
the blue and red GCs from the ACS surveys. This shows the reliability of 
our transformations.

The nucleus $V$ \& $(V-I)$ values for Fornax dwarf ellipticals were taken
from Lotz et al.\ (2004). For the Fornax UCDs, we present our ACS sample 
plus additional objects from Gregg et al.\ (2008) 
with colors from Karick et al.\ (2008): 
the original Sloan $gri$ magnitudes were transformed to $VI$ via the 
transformation equations of 
Lupton (2005)\footnote{See http://www.sdss.org/dr4/algorithms/sdssUBVRITransform.html.}.

Looking at the CMDs, we observe that the UCDs and ``dwarf globular transition 
objects'' are spread over the same color range as GCs. The
brightest UCDs ($M_V<-12.3$ mag), however, seem to favour red colors, thus probably are an
extension of the metal-rich GC peak. Interestingly, Wehner \& Harris (2007)
found such an extension of red GCs towards the UCD regime in the
extraordinary rich globular cluster system of NGC3311, the central
galaxy in the Hydra\,I Cluster.

The apparent ``gap'' in luminosity for the objects around $M_V\sim-10$ mag is a
pure selection and incompleteness effect. On the one hand, the GCs from the
ACS surveys only cover a small area around galaxies and are selected by their apparent
sizes and magnitudes. UCDs would have been rejected by these surveys as extended background galaxies. 
On the other hand, the redshift-selected UCDs cover
a large area in the clusters but suffer from the completeness limit of
spectroscopic surveys. Our Fornax Cluster survey has a magnitude limit
of $M_V\sim-10.3$ mag (roughly)\footnote{We note that some UCDs in
  Figure 5 have lower luminosities than the nominal survey limit. This
is because the original survey selection was based on photographic photometry with
relatively high uncertainties.}. 

An additional feature is visible in the Fornax CMD. There is a group
of UCDs (with $M_V\sim-10.5$ mag) which exhibit  very blue colors,
$V-I\leq0.9$ mag. Such colors can be interpreted in two ways: either these
UCDs are very metal-poor ([Fe/H]$\sim-2$ dex), or their stellar populations are 
of intermediate-age. Indeed, some UCDs in Fornax show increased H$\beta$ line
indices (Mieske et al.\ 2006), a hint to the contribution of young stellar
populations.

In the Virgo CMD, we plot the color-magnitude relation for dwarf galaxy nuclei
(host galaxy magnitude $B_T>13.5$ mag) from the ACS Virgo Survey
(C\^ot\'e et al.\ 2006). A large fraction of UCDs (as well as ``dwarf globular 
transition objects'') falls close to this relation. 
Interestingly, the nuclei of early-type giant galaxies 
cluster around red colors, $V-I\sim1.25$ mag, at the same location
where the brightest and also the reddest UCDs are found. 
We can not distinguish UCDs from nuclei of early-type galaxies by simply 
using magnitudes and colors.

In the Fornax CMD, we also show nuclear star clusters (NCs) of late-type galaxies, although  
the galaxies are not Fornax members. The data are from the work of Rossa et al.\ (2006), 
based on previous works by Walcher et al.\ (2005, 2006) and B\"oker et al.\ (2002, 2004). 
The data were corrected for the foreground extinction but the intrinsic
extinction of the host galaxy is more problematic. Only few NCs in the Rossa et al.\ and B\"oeker et al.\ lists 
have estimates of the internal reddening (and were corrected for it).
As we mentioned in the previous section, NCs of late-type galaxies and nuclei of early-type galaxies 
are nearly identical in the sense of luminosities and sizes. The only difference is that 
the majority of NCs have young ages, they are younger than early-type galaxy nuclei. 
Looking at Figure~5, one can see that there are some blue (young) NCs that are close to the
location of blue GCs and UCDs or at least should pass this location when
aging. There are quite a few NCs with very blue colors ($V-I<0.7$ mag) and bright magnitudes 
($M_V<-10$ mag), which can not be seen in Figure~5 due to the scale of the figure. 
There are also many red NCs that probably are contaminated
by internal reddening (so they might actually be bluer). 
The blue NCs look very bright, on average brighter than early-type galaxy nuclei and UCDs.
The bright magnitudes can be caused by young ages. Younger stellar
populations are not only bluer but also much brighter. They will fade with time. Red bright NCs might
be intrinsically very luminous/massive. They come mostly from MW-type
spirals. As above (in Section~4.1), the main conclusion we can draw out of 
the comparison of UCDs with NCs is that 
some of the UCDs, both blue and red, could be threshed nuclei of late-type galaxies.

\subsection{Color profiles and color gradients}

The analysis of radial color gradients for the UCDs is very challenging, mainly 
because they are tiny and very compact objects, hardly resolved even with the HST. 
So the corrections for telescope PSF effects become extremely important. 
In this section we therefore reanalyse the images
to obtain the best possible estimates of the radial color profiles and
devise a series of tests to quantify how the PSF and other issues affect the color profiles.
Note that we present results for the brightest UCDs only (Virgo UCDs and Fornax UCD3 \& UCD6) 
as the color gradients for the fainter UCDs could not be estimated with any reliability.

\subsubsection{Measurement of color gradients}

We obtain the radial color profiles by fitting UCD images 
with PSF-convolved models using {\sc GALFIT}. Superficially 
it may appear easier to derive color profiles directly from PSF-deconvolved images. 
However, we choose not to do this,  
as a deconvolution process (e.g.\ the Lucy-Richardson algorithm) amplifies 
noise in the image and, in our case, produces very large artificial fluctuations in the color profiles. 
Instead we derive color profiles from {\sc GALFIT} models. 
We do not use the models obtained in Section~3 as these were designed to obtain the best parameters 
for standard models where possible. The differences to the approach in Section~3 are as follows: 
\begin{enumerate}
\item Multiple component models (e.g.\ King+S\'{e}rsic+S\'{e}rsic) are used to better fit the UCDs. 
No constraints that the model parameters must have physically meaningful values are applied.  Our aim
is to match the UCD surface brightness (SB) profiles as well as possible.
Multiple S\'{e}rsic models are the best for almost all the UCDs.
\item The PSFs are presented out to a radius of $10''$ in the modeling.  
We also do several experiments with the PSFs to make sure that the color gradients 
we find for UCDs are robust.
\item We fit the background directly rather than holding it fixed.
\end{enumerate}

We start by using {\sc GALFIT} to fit more general multi-component models
to each UCD and filter combination.
From the best-fitting parameters we generate the two-dimensional model 
(using {\sc GALFIT} again), which is in principle free of PSF effects.   
Then we calculate SB profiles for the models in each filter using the {\sc IRAF} task {\sc ELLIPSE}. 
The color profile is then determined as the difference of the SB profiles in the two filters. 
The resulting color profiles are presented in the top panels of Figures~7--16. 

In the rest of this subsection we 
discuss the most important issues which affect the color profiles, such as  
accuracy of the {\sc GALFIT} model fits to the data, PSF effects and sky subtraction.

{\it 1.  Accuracy of the {\sc GALFIT} model fits to the data}

The object-minus-model residuals are shown in the bottom panels of Figures~7--16. The residuals 
get larger with radius. This is normal and expected. 
If we plot the error bars on the UCD surface brightness profiles produced by {\sc ELLIPSE} 
(errors on measuring the mean flux along each isophote), we find that they are are comparable 
to the size of the residuals. The most important point here is to fit the objects so that 
the residuals do not show any systematic trend (so that the residuals fluctuate evenly around zero). 
In some cases, the residual fluctuations are larger than the {\sc ELLIPSE} errors, but this could be 
because the {\sc ELLIPSE} error bars are only statistical and do not include any other possible errors.

{\it 2.  PSF effects}

The PSF depends on the object color. In the same broad-band filter, the PSF (PSF FWHM) of a red star may be 
noticeably larger than that of a more blue star. 
We therefore generated three PSFs using {\sc TinyTim} (and {\sc MultiDrizzle}, as described in Section~3): 
one with the average UCD color, one with the reddest and one with the bluest possible with {\sc TinyTim} color. 

The PSFs were made of a very large size ($20''\times20''$ or $800 \times 800$ pixel, larger than the UCDs) 
to attempt to address the PSF halo problem. The ACS/HRC chip has a defect that creates a halo surrounding the
PSF at wavelengths $> 0.6 \mu$m and the relative proportion of flux within this
halo increases with wavelength. The halo is large (many arc seconds) and can contain 10-20\% of the total flux.
{\sc TinyTim} models this effect, but unfortunately not very well\footnote{See http://www.stsci.edu/software/tinytim.}. 

Stellar images (``real'' PSFs) are better representations of the {\it true} PSF, but they also have some disadvantages.
All the stellar images we found in the HST archive suitable for the UCD 
image modeling (i.e.\ images of a single star which was centrally located and non-saturated) are much bluer 
than the UCDs. 
Another disadvantage of these PSFs is that they have a very low S/N at large radii and, therefore, can not correctly 
capture the outer PSF halo. 
Brighter stars, which have higher S/N at large radii, are saturated in the center and, hence, can not be used for modeling, 
but we can use them for the comparison with the {\sc TinyTim} PSFs,  
to check how reliable the {\sc TinyTim} PSFs are at large radii. 
We have managed to find in the archive a couple of high S/N stars with red colors, similar to 
UCD colors.     

In Figure~6 we plot the color profiles for {\sc TinyTim} PSFs, a low S/N star (with blue color) and a high S/N star 
(with red color), and normalize all of them at $\sim 6$ pixels.
All {\sc TinyTim} PSFs look very similar (except at very large radii, $R > 100$ pixels) and give very similar results 
for the UCD color profiles in Figures~7 \& 8. So for the UCD image modeling, we can safely use just one {\sc TinyTim} PSF 
with the average UCD color. 
Another important conclusion we can draw from Figure~6 is that the color profiles of the high S/N star 
(with red color similar to UCD colors) and {\sc TinyTim} PSFs are nearly the same in the region $\sim 4 - 54$ pixels. 
The match between the color profiles of the low S/N star (with blue color) and {\sc TinyTim} PSFs is a bit worse in the same
radius range (compared to the high S/N PSF). 

In Figures~7--16 we present the UCD color profiles, obtained with the {\sc TinyTim} PSF (with 
the average UCD color and of a very large size) and the low S/N star, but due to the above reasons we prefer to 
trust the {\sc TinyTim} PSF more.

Having done all of this, we conclude that the color
gradient is most reliable in the region $\sim 4 - 54$ pixels ($\sim 7.5 - 100$ pc 
for Virgo and $\sim 9 - 124$ pc for Fornax). The uncertainties in the 
PSF structure become significant outside this radius range.  
In addition to the PSF effects, we can not trust the color gradient beyond $R \sim 60 - 80$ pc 
for most of the UCDs due to the noticeable deviation between
the {\sc GALFIT} model and the data. The color gradient may be incorrectly amplified because of it. 
To highlight the region where we consider the color gradient is reliable (the {\it ``trusted region''}), 
we draw two vertical lines in all the panels:  
at the outer radius where object-minus-model residuals start to deviate from zero significantly,   
and at the inner radius where there may be uncertainties in the structure of the PSF core.  For 
VUCD3, for example, the ``trusted region'' is $\sim 7.5 - 75$ pc (Figure~7).

{\it 3.  Sky subtraction} 

The sky (background) was initially subtracted from the UCD images by {\sc MultiDrizzle}. 
We then applied additional background corrections, determined from empty parts of the images.
Finally, we fitted the background when fitting UCDs with {\sc GALFIT}. 
For the latter, we did the fitting over very large area, masking out the HRC ``occulting finger'' and 
all the background/foreground objects in the image, except objects appearing 
very close to the UCDs. We fitted and subtracted such objects from the image, which is more accurate than masking. 
Note that besides the mean sky level, {\sc GALFIT} finds the gradient in the sky.  
The sky level is determined very well, as seen from the absence of a trend in the residuals in 
Figures~7--16 (except for VUCD2). \\

The results of the color profile fitting are shown in Figures~7--16. The main observational conclusion
is that, with two exceptions, all the UCDs appear to have small positive color gradients in the sense 
of getting redder outwards: 
mean $\Delta(F606W-F814W) = 0.14$ mag per 100 pc with $rms = 0.06$ mag
per 100 pc. The two exceptions are VUCD2 and UCD3.
It is hard to make very strong conclusions regarding VUCD2 because of the large fluctuations in the color 
profile and a slight systematic trend (upwards) in the residuals.
We would also consider the UCD3 color gradient as unreliable, since 
we did not model and subtract the background spiral (see Section~3 and Evstigneeva et al.\ 2007). 
It is very hard to separate this faint object from the UCD reliably. 
We did all the modeling for the half of the UCD, which is less affected by the spiral, but we can not 
prove that it is completely unaffected. There still can be some flux from this spiral, projected onto the UCD3 center
and making it slightly bluer.

\subsubsection{Interpretation}

We stress that this discussion only applies to the brightest UCDs for
which it was possible to estimate the color gradients.

First we consider the hypothesis that UCDs are very luminous GCs.

Djorgovski et al.\ (1991) report color gradients in MW GCs. They find
that the clusters with post core collapse morphology (with central
cusps) have positive color gradients (in the same sense as the
gradient in the UCDs), while clusters with King model morphology (with
flat cores) do not show any color gradient.  Djorgovski et
al.\ interpret this as evidence that the dynamical evolution of
clusters can modify their stellar populations.  Core collapse tends to
affect more massive stars, with lower mass stars (red sub dwarfs)
tending to move to larger average radii. The Djorgovski et al.\ study
may not, however, be relevant to UCDs. Their measurements 
only extend to 1--8 pc from the cluster center and, more
importantly, relaxation and mass segregation (less massive red
sub dwarfs sitting outside) would not work for systems as large as
UCDs. Their half-mass relaxation times are $\sim 8$ Hubble times (for
the Virgo UCDs with the dynamical and structural parameters from
Evstigneeva et al.\ 2007) and larger than a Hubble time for radii
$R>8$\,pc where we have our color gradient measurements.  There has
not been sufficient
time to drive the low mass stars towards large radii.  Also, in a
later work, Sohn et al.\ (1996) found positive color
gradients in GCs with King model morphology (they went to a larger
radius from the center, 11 to 19 pc), and made a conclusion that the
color gradient is not unique to post core collapse GCs and, hence,
may not be explained by the dynamical evolution. Still more recently,
Sohn et al. (1998) found both negative and positive color gradients
in GCs with both post core collapse and King model morphology.

Other possible interpretations of the UCD color gradients could be that  
UCDs were born mass-segregated (primordial mass segregation, if it has not been erased 
because the relaxation times are long); or that UCDs are
composed of multiple (at least two) stellar populations. 
It is actually unlikely we are seeing mass segregation as most of the
light from these old populations is mainly coming from upper main sequence stars 
and giants with a relatively small mass range. On the other hand, some of
the brightest MW GCs (e.g.\ $\omega$~Cen, NGC2808, NGC1851) are
found to contain multiple stellar populations (e.g.\ Milone et
al.\ 2008).  $\omega$~Cen is the only GC for which we have information
about the radial distribution of stellar populations (Hilker \&
Richtler 2000, Sollima et al.\ 2007).  Hilker \& Richtler (2000)
studied the two major populations (of RGB stars) in $\omega$~Cen (it has at least three
sub-populations) and found that the younger, more metal-rich population is more
centrally concentrated than the older, more metal-poor population. This
gives a negative color gradient---opposite to what we find for UCDs.
A possible explanation for $\omega$~Cen is that a more centrally concentrated
population formed at a later time from enriched material.
Assuming the same scenario for UCDs, we can test if it agrees with the
UCD color
gradients.  For Virgo UCD ages ($8 -15$ Gyr), metallicities
([Z/H]=$-1.35$ to $-0.17$ dex) and colors ($V-I=0.96-1.13$ mag) (Evstigneeva
et al.\ 2007) and using Maraston (2005) SSP models, we consider the
following model populations: a more metal rich ([Z/H]=-0.33 dex), younger
(8 Gyr) population with $V-I=1.08$ mag, and a more metal poor ([Z/H]=-1.35
dex), older (15 Gyr) population with $V-I=0.98$ mag. The colors are obtained 
for a Salpeter IMF (the Kroupa IMF
gives very similar results). The sum of these two populations will give
total ages, metallicities and colors similar to those observed in the
UCDs. The color gradient, however, is
negative---opposite to what is observed in the UCDs (although the same as in $\omega$
Cen). This suggests that the star-formation histories in UCDs and $\omega$
Cen are different, but this still does not prove that UCDs are different to
GCs.  Recent work by Milone et al.\ (2008), for example, says that
three different GCs with multiple stellar populations ($\omega$~Cen,
NGC2808, NGC1851) all have different star-formation histories.  Milone
et al.\ also note that  ``the star-formation history of a GC can vary
strongly from cluster to cluster'' and they point out that the multiple
stellar populations have only been detected in the most massive MW GCs. 
They suggest that ``cluster
mass might have a relevant role in the star-formation history of GCs.''

From these results we conclude that UCDs may contain multiple stellar
populations without contradicting the hypothesis that they are bright
GCs. However a lot more has to be done in terms of both observational
work and theory/simulations (as was concluded by D'Antona \& Caloi
2007) to understand the origin of multiple stellar populations in UCDs
and GCs. Obviously, something happens in very massive/luminous GCs so
that they have multiple stellar populations.  It might be that their
gravitational potential is strong enough to keep the gas that would
otherwise be completely expelled by stellar winds and supernovae
explosions (see e.g. Baumgardt et al.\ 2008) and that the second
(and other) generations of stars formed from this gas.

We now consider the alternative hypothesis, that UCDs are the nuclei
of threshed galaxies. 

It could be argued that the color
gradients in UCDs are consistent with the threshing hypothesis for the
formation of UCDs from dE,Ns.  Lotz et al.\ (2004)
showed for Fornax cluster dE,Ns, using aperture photometry, that their
stellar envelopes are 0.1-0.2 mag redder in $V-I$ than their nuclei.
In this case, residual stars from the envelope after the disruption process would
naturally explain the color gradient in UCDs.  We can test this with
more recent data from the ACS Virgo Cluster Survey for early-type
galaxies. We compared the colors of the nuclei (C\^ot\'e et al.\ 2006) and
their underlying galaxies (Ferrarese et al.\ 2006): the nuclei are
bluer than their underlying galaxies for dEs ($M_B>-17.6$ mag), which is
consistent with the Lotz et al.\ findings, while for giant Es
($M_B<-17.6$ mag) the nuclei are redder than their underlying galaxies.
These results are in agreement with the color gradients found in
dwarf and giant elliptical galaxies. Dwarf elliptical galaxies have
mostly positive color gradients (getting redder outwards, e.g. Vader
et al. 1988), while luminous early-type galaxies have generally negative
color gradients (getting bluer outwards, e.g. de Propris et
al.\ 2005a). Vader et al.\ (1988) explain the origin of
positive color gradients in early-type dwarf galaxies by age
gradients, as metallicity gradients are removed by galactic winds.
Brighter galaxies have a deeper potential
well. For them, galactic winds will be important in the outer parts
only. This is why bright galaxies have stronger metallicity gradients.

If UCDs are formed by {\em simple} stripping of dEs (with no other
processes involved, like effects of gas removal on the chemical
evolution of nuclei), the UCD color gradients are in agreement with
the threshing hypothesis (with both dE color gradients and dE
core/envelope colors). There is a problem with the brightest and
reddest UCDs, however.  If we have a look at the CMDs (Figures~4 \& 5),
VUCD3, VUCD7 and UCD3 are all located in the region where only giant
elliptical galaxy nuclei are found.  If we interpret the fact that
VUCD7 and UCD3 have envelopes as incomplete stripping and consider
only the cores of these UCDs in the CMDs, the cores will then be in
the same region as dE nuclei. (The cores of all the UCDs with
envelopes have magnitudes and colors as follows: $-11.16 \geq
M_V\geq-12.39\,{\rm mag}$, $0.92 \leq V-I\leq 1.18\,{\rm mag}$.) Thus,
VUCD7 and UCD3 may still be threshed dE galaxies.  VUCD3 does not have
an envelope but it is very bright and very red, so it does look like a
giant elliptical galaxy nucleus. However, VUCD3 has a positive color
gradient which we might not expect in a giant elliptical galaxy.  It
may be possible to argue that VUCD3 is a threshed late-type galaxy (if
we still prefer the threshing hypothesis), since late-type galaxies
have mostly positive color gradients (e.g. Taylor et al.\ 2005).

It may also be possible that UCDs are genuine, but anomalous, dwarf galaxies
(requiring no threshing). Their positive color gradients are
consistent with dE galaxy color gradients---the pure age gradients,
explained by galactic winds (as above).  If we assume that UCDs have
younger stellar populations in the center (7 Gyr) and older populations
in the outer regions (15~Gyr), then for a metallicity of [Z/H]=-0.33
dex (same for the two populations), Maraston (2005) SSP models with a
Salpeter IMF will give us the color variation from $V-I=1.06$ mag in the
center to $V-I=1.16$ mag outside; or for a metallicity of [Z/H]=-1.35 dex:
from $V-I=0.90$ mag in the center to $V-I=0.99$ mag outside. This is
consistent with what we see in the UCDs. 

Our conclusion is that the positive color gradients found in bright
UCDs (2 Fornax UCDs and 8 Virgo UCDs) are consistent with them being
either bright GCs or threshed dE galaxies (except for VUCD3).  However,
the spectroscopic ages, metallicities, and $\alpha$-abundances for
Virgo UCDs, obtained in our previous work (Evstigneeva et al. 2007),
are not consistent with the formation of UCDs by the simple removal of
the envelope from the nuclei of dE galaxies. Since spectroscopic ages and
metallicities are more powerful tools than colors (colors are
degenerate), we have to reject the threshing hypothesis for the Virgo
UCD origin. Hence, the Virgo UCDs are more consistent with being
bright GCs. As for the two Fornax UCDs, no firm conclusions on their
origins can be drawn without having spectroscopic age, metallicity,
and $\alpha$-abundance estimates for them. At the moment, we can only
say that their color gradients are consistent with all three
hypotheses: threshed dE galaxies, bright GCs, and genuine dwarf galaxies.

\section{Summary}

In this paper we have presented analysis of the structure and colors
of the most extensive and complete sample of UCDs in the Fornax and
Virgo Clusters observed to date: 13 Fornax UCDs with magnitudes in the
range $-10.49 \geq M_V\geq-13.33\,{\rm mag}$ and 8 Virgo UCDs with
magnitudes $-11.95 \geq M_V\geq-13.42\,{\rm mag}$.  The sample
includes 6 Virgo UCDs initially presented by Evstigneeva et al.\
(2007), as we have made several improvements to the image analysis.
The main results of our analysis are as follows:

\begin{enumerate}
\item We have modelled the images of Fornax and Virgo UCDs using the
  two-dimensional fitting algorithm {\sc GALFIT}, assuming
  empirical King, S\'{e}rsic and Nuker models (or King+S\'{e}rsic and
  S\'{e}rsic+S\'{e}rsic) for the luminosity profile. We find that for
  the faint UCDs ($m_{V,0} \sim 19.9 - 20.9$ mag), there is no preference of
  one model over another: all the models (Nuker, King and S\'{e}rsic) fit
  the UCDs equally well.  The bright UCDs ($m_{V,0} \sim 17.5 - 18.9$ mag)
  have extended outer halos best fitted by two-component
  models (King+S\'{e}rsic or S\'{e}rsic+S\'{e}rsic) or a double power law (Nuker).
  This does not mean that bright and faint UCDs are intrinsically
  different.  With deeper observations for the fainter UCDs, we
  could possibly detect outer halos in them were any present.  The
  detection of these extended halos around luminous UCDs is not
  consistent with their having the standard King profiles
  traditionally associated with GCs, but recent work of McLaughlin \&
  van der Marel (2005) and McLaughlin et al.\ (2008) has shown that extended halos are 
  a general characteristic of massive GCs in the Milky Way and some of its satellites 
  and NGC5128.

  Our modeling also shows that the UCDs have a range of 
  S\'{e}rsic indices $n$ and King concentrations $c$, as well as 
  a range of central slopes (seen from Nuker inner slopes $\gamma$): 
  from flat ``King'' cores to central cusps. This suggests that 
  convolving some cuspy models with the PSF allows such models 
  to fit the seeing-blurred centers of some UCD profiles. 
  We, however, can not make very strong conclusions regarding central 
  cusps or cores in the UCDs, taking into account the resolution and 
  other limits of our data.  
 
  We do not find any correlations between the model parameters mentioned above  
  and the UCD luminosity, size or color. Furthermore, no correlation
  of ellipticity with luminosity, size or color has been found for
  the UCDs.

  The two-sample K-S and Wilcoxon tests show that the UCD
  ellipticities are consistent with extragalactic GC distributions
  (NGC5128 and M31 GCs), but significantly different from the MW GC
  distribution.

\item We have shown that UCDs and GCs form a continuous distribution
  across the luminosity-size plane, but UCDs seem to be different to
  {\em typical} GCs in the sense that the UCDs have sizes correlated
  with luminosities whereas the GCs do not.  We obtain the following
  luminosity-size relation for the UCDs in our sample: 
  $R_{\rm eff} \propto L_V^{0.88 \pm 0.13}$ or $R_{\rm eff} \propto L_V^{0.68 \pm 0.13}$ 
  (if we exclude two brightest objects).  However, the
  luminosity-size relation we observe for the UCDs is similar to
  observations for most {\em luminous} GCs: Barmby et al.\ (2007)
  report an increasing lower bound on $R_{\rm eff}$ in the mass vs.\
  $R_{\rm eff}$ plane for the most massive GCs.

  The nuclei of early-type galaxies (from C\^ot\'e et al.\ 2006) 
  with luminosities similar to the UCDs show a
  luminosity-size correlation with the same slope as the UCD relation,
  but we find that the effective radii of the UCDs are
  systematically $\sim 2.2$ times larger than the nuclei of the same
  luminosity. This difference is significant at a confidence level
  $>99.9\%$.

  Although this observation is superficially inconsistent with a
  process whereby these galaxy nuclei are stripped to form UCDs, the
  change in size may be a result of the stripping process.  The
  simulations of the stripping process by Bekki et al.\ (2001, 2003)
  indicate some expansion of the remnant core, but no quantitative
  results are provided. Given our new observational result, it will be
  important to test this in detail against simulations of the
  threshing process to see if it is consistent with the size
  difference we have found between UCDs and the early-type galaxy
  nuclei.

  We also note that the nuclei of late-type galaxies are similar to
  the nuclei of early-type galaxies (and UCDs) in terms of
  luminosities and sizes.  This means that UCDs---if formed by
  disruption---could be the remnant nuclei of both early-type and
  late-type galaxies, not just dE,Ns as it was initially suggested.

\item The UCD total magnitudes and colors (their position in the
  color-magnitude diagrams) are consistent with them being either
  luminous GCs or threshed nuclei of both early-type and late-type
  galaxies.

\item We have estimated radial color gradients for the brightest UCDs
  in our sample: Virgo UCDs and Fornax UCD3 \& UCD6.  All the UCDs
  (with two exceptions) appear to have small positive color gradients
  in the sense of getting redder outwards: 
  mean $\Delta(F606W-F814W) = 0.14$ mag per 100 pc with $rms = 0.06$ mag per 100 pc.  
  The two exceptions are VUCD2 and
  UCD3: the color gradient estimates for them are unreliable.

  The positive color gradients found in the bright UCDs are consistent
  with them being either bright GCs or threshed dE galaxies (except
  VUCD3). However, the spectroscopic ages, metallicities, and
  $\alpha$-abundances for Virgo UCDs, obtained in our previous work
  (Evstigneeva et al. 2007), are not consistent with the formation of
  UCDs by the {\em simple} removal of the envelope from the nuclei of
  dE galaxies. Since spectroscopic ages and metallicities are more
  powerful tools than colors, we have to reject the threshing
  hypothesis for the Virgo UCD origin.  Hence, the Virgo UCDs are more
  consistent with being bright GCs. As for the two Fornax UCDs, no
  firm conclusions on their origins can be drawn without having
  spectroscopic age, metallicity, and $\alpha$-abundance estimates for
  them. At the moment, we can only say that their color gradients are
  consistent with them being threshed dE galaxies, bright GCs, and
  genuine dwarf galaxies.
\end{enumerate}

The aim of our investigation was to test two formation hypotheses for
UCDs---whether they are bright GCs or threshed early-type dwarf
galaxies---by direct comparison of UCD structural parameters and
colors with GCs and galaxy nuclei. In most of the
measurements we have made (profiles; color-magnitude relations;
color gradients), the UCDs display properties consistent with a
threshing origin and with what might be expected for luminous GCs. 
We therefore conclude that these structural parameters and colors 
are not able to distinguish sufficiently between the different 
formation hypotheses.

The one exception to this conclusion is the difference we find in
the size-luminosity relation between UCDs and the nuclei of early-type
galaxies. This significant difference (the UCDs are 2.2 times as large
as nuclei at the same luminosity) suggests an important numerical test
of the threshing hypothesis: it should be relatively easy to predict
the increase in the size of nuclei resulting from the threshing
hypothesis and compare that to our new observational results.

\acknowledgments

This work has been supported by a Discovery Project grant from the Australian 
Research Council.
Part of the work reported here was done at the Institute of Geophysics and 
Planetary Physics, under the auspices of the U.S. Department of Energy by 
Lawrence Livermore National Laboratory in part under Contract  W-7405-Eng-48 
and in part under Contract DE-AC52-07NA27344.
The authors wish to thank Harry Ferguson (STScI) and Anton
Koekemoer (STScI) for their assistance with the HST data analysis, 
Holger Baumgardt (University of Bonn) for helpful discussions and the referee 
for their very careful reading of our paper and many helpful suggestions.



{\it Facilities:} \facility{HST (ACS)}




\clearpage



\begin{figure}
\epsscale{1.05}
\plotone{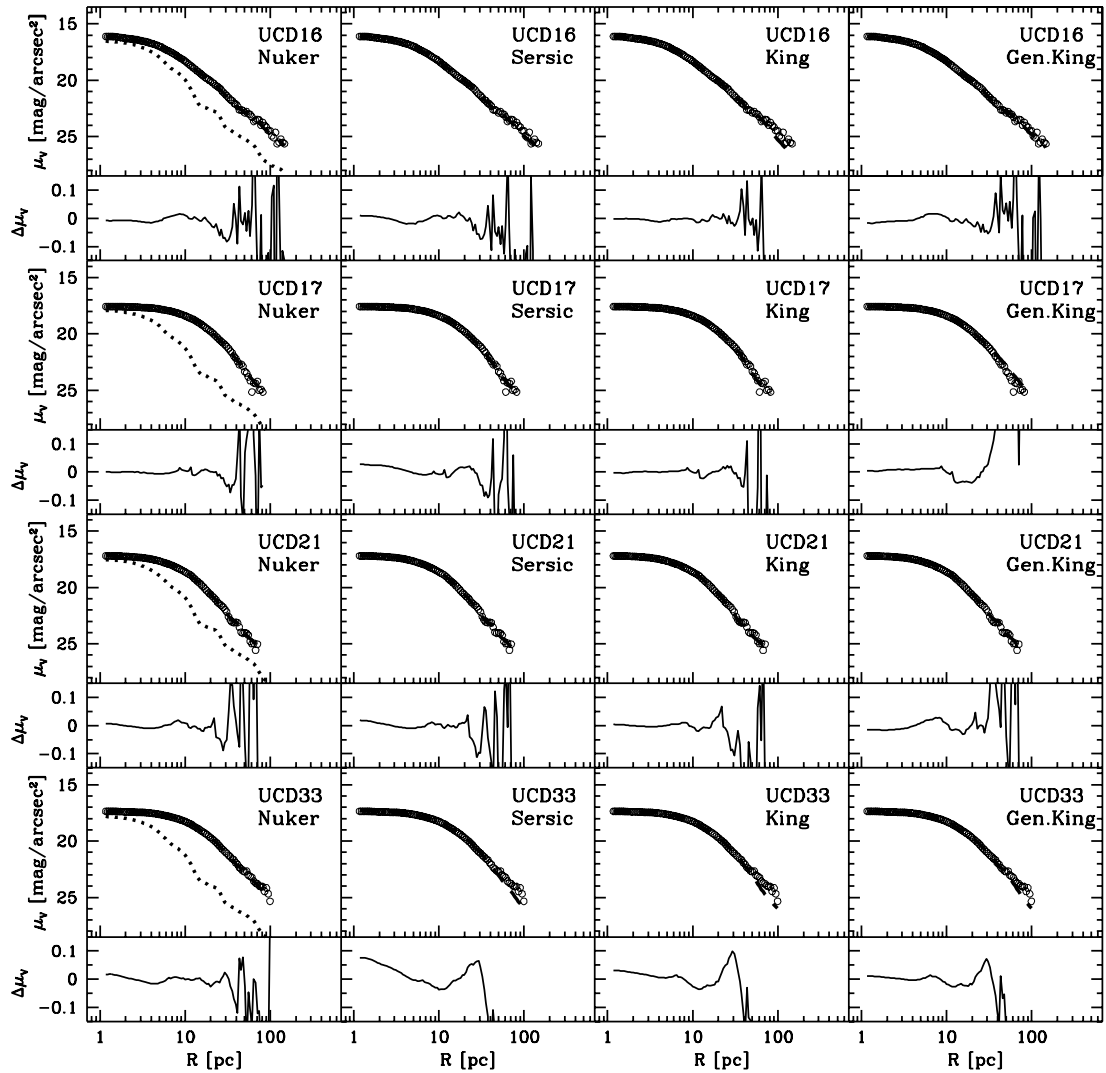}
\caption{Surface brightness profiles, measured in the F606W images, and model fits to 4 UCDs. 
The plots for the other 17 UCDs are available in the electronic edition of the Journal (Figures~1.1.--1.5). 
The magnitudes are AB magnitudes. 
The open circles represent the UCD profile, the dashed line represents the best-fitting model, convolved 
with the PSF. The PSF for each object is shown with the dotted line. $\Delta \mu_V$ plots show the residuals 
for each fit: the difference (in magnitudes) between the UCD profile and the profile of the best-fitting model, 
convolved with the PSF.}
\end{figure}






\begin{figure}
\epsscale{.90}
\plotone{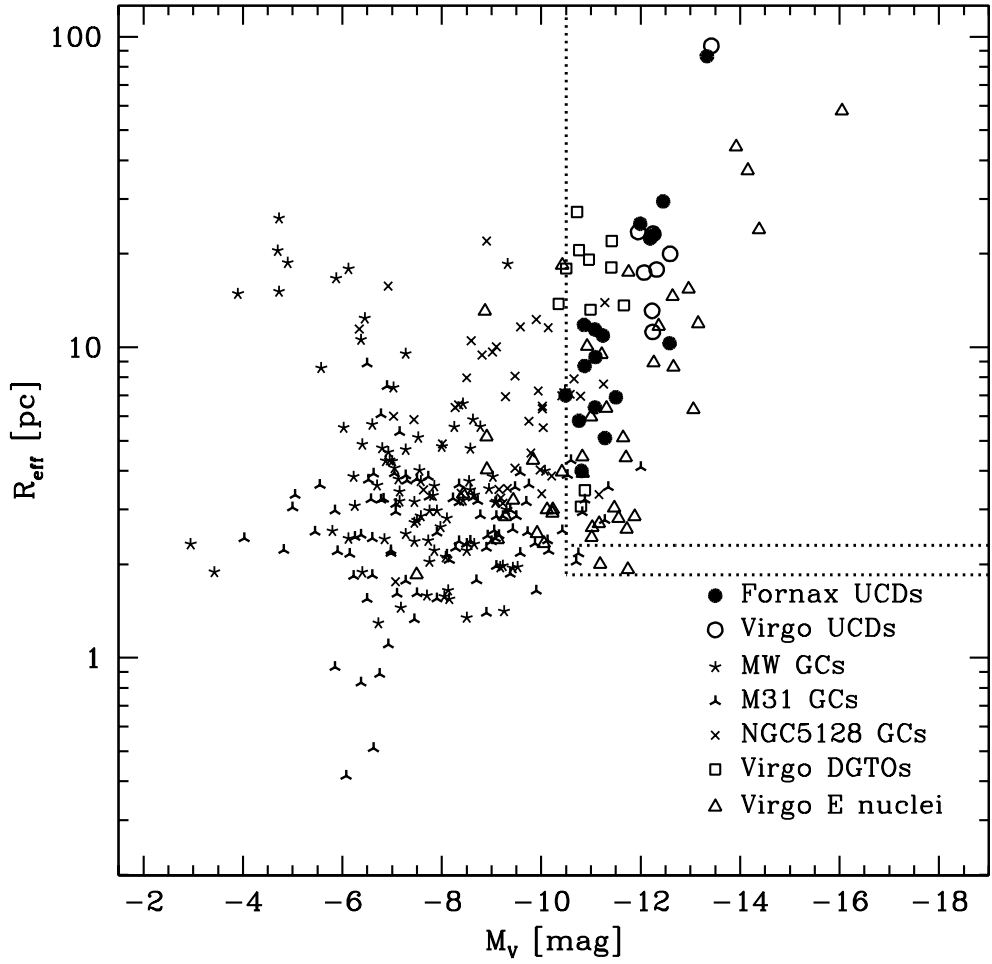}
\caption{Luminosity -- Size diagram.  
Fornax and Virgo UCDs: this work (Table~1). 
MW GCs: McLaughlin \& van der Marel (2005), photometry is based on Wilson models.
M31 GCs: Barmby et al.\ (2007), photometry is based on King models.
NGC5128 GCs: Holland et al.\ (1999), Harris et al.\ (2002).
Virgo DGTOs: Ha\c{s}egan et al.\ (2005), certain DGTO candidates, 
$R_{\rm eff}$ is the mean of the two pass bands, $g'$ and $z'$. 
Virgo early-type galaxy nuclei: C\^{o}t\'{e} et al.\ (2006), all resolved nuclei, 
$R_{\rm eff}$ is the mean of the two pass bands, $g'$ and $z'$. 
The vertical dotted line shows the magnitude limit of our 2dF/HST UCD observations. 
The horizontal lines show the HST/ACS resolution limit: the lower line is 
for the Virgo distance and upper line is for Fornax (below these lines $R_{\rm eff}<1$ pix).}
\end{figure}

\begin{figure}
\epsscale{.90}
\plotone{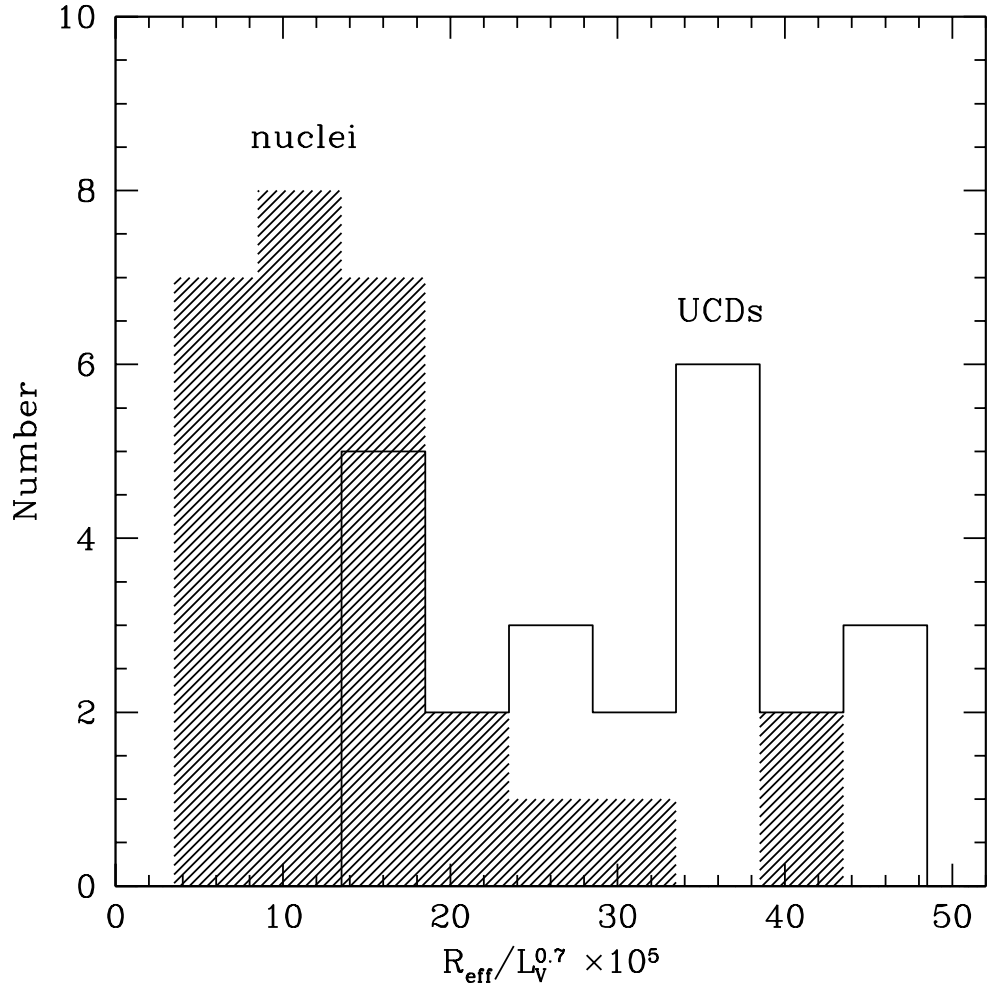}
\caption{Size comparison of UCDs and early-type galaxy nuclei. The
  sizes are all scaled relative to the size-luminosity relation fitted
for the nuclei in Equation 9.}
\end{figure}

\begin{figure}
\plotone{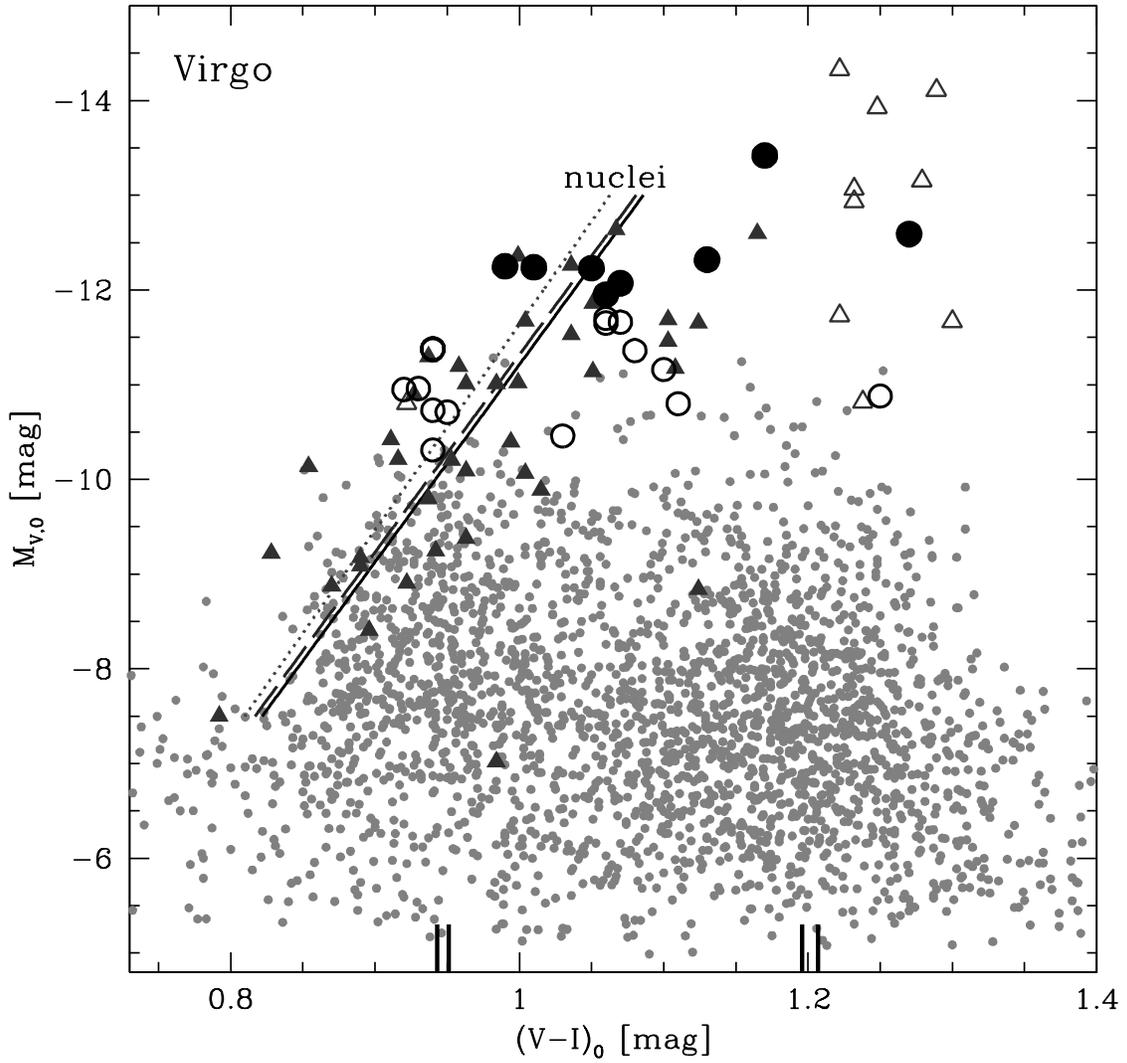}
\epsscale{.90}
\caption{Color-magnitude diagram for the Virgo Cluster objects. 
Filled circles are Virgo UCDs from the present work. Open circles are ``certain and probable'' 
Virgo ``dwarf globular transition objects'' from Ha\c{s}egan (2005).   
Grey dots are M87 and M49 GCs from the ACS Virgo Cluster Survey. Triangles are nuclei of 
early-type galaxies from the ACS Virgo Survey: 
open triangles -- nuclei of galaxies brighter than $B_T=13.5$ (approximate dividing point 
between dwarf and giant galaxies), filled
triangles -- nuclei of galaxies fainter than $B_T=13.5$.  
The line is the fit to the sample of nuclei of galaxies fainter than $B_T=13.5$ (three lines 
correspond to three different age bins used for magnitude transformations: 
solid line -- 11-13 Gyr, dashed line -- 6-8 Gyr, dotted line -- 2-4 Gyr). 
Ticks on the bottom: peak colors for the blue and red
GC population of M87 and M49 from Larsen et al.\ (2001).}
\end{figure}

\begin{figure}
\epsscale{.90}
\plotone{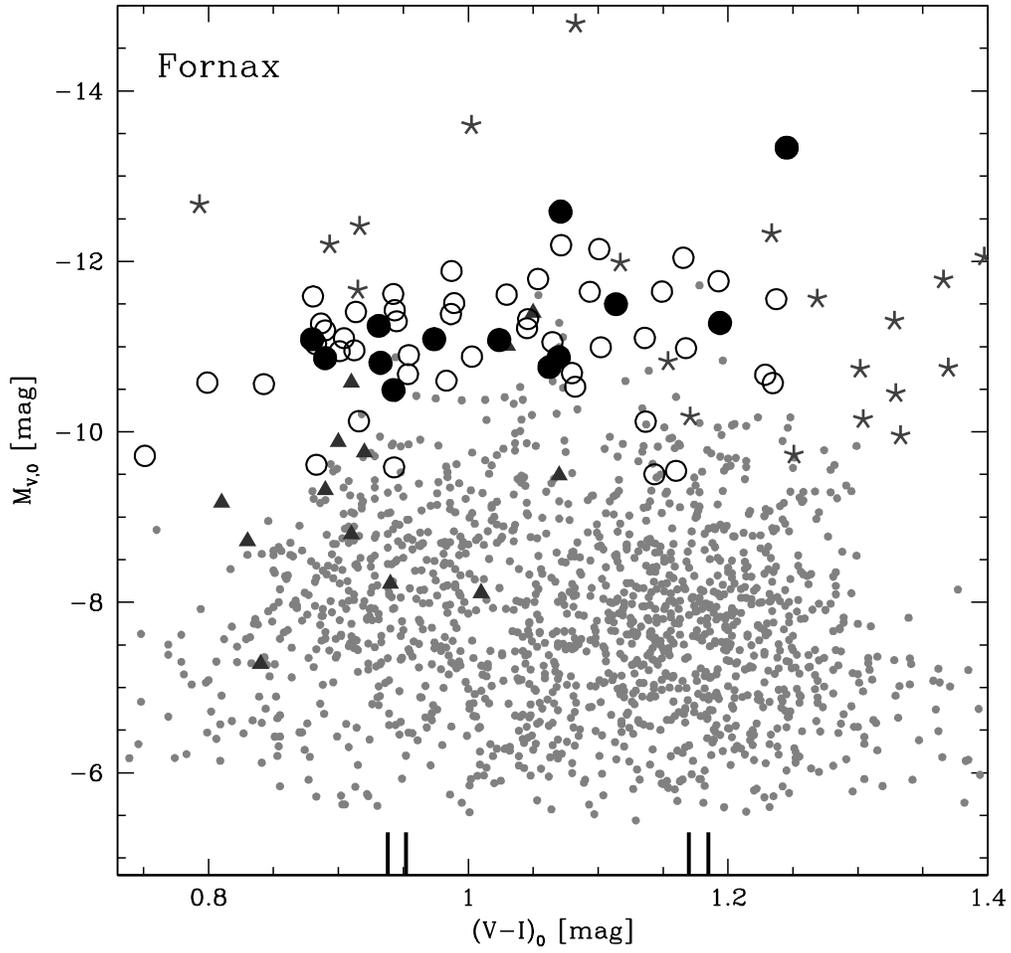}
\caption{Color-magnitude diagram for the Fornax Cluster objects.  
Filled circles are Fornax UCDs from the present work.  
Open circles are Fornax UCDs from Gregg et al.\ (2008).  
Grey dots are NGC1399 and NGC1404 GCs from the ACS Fornax Cluster 
Survey. Filled triangles are dwarf elliptical nuclei from Lotz et al.\ (2004). 
Ticks are for the blue and red GC peaks of NGC1399 and NGC1404 (Larsen et al.\ 2001). 
Asterisks are nuclear star clusters (NCs) of late-type galaxies (Rossa et al.\ 2006).}
\end{figure}

\begin{figure}
\epsscale{.95}
\plotone{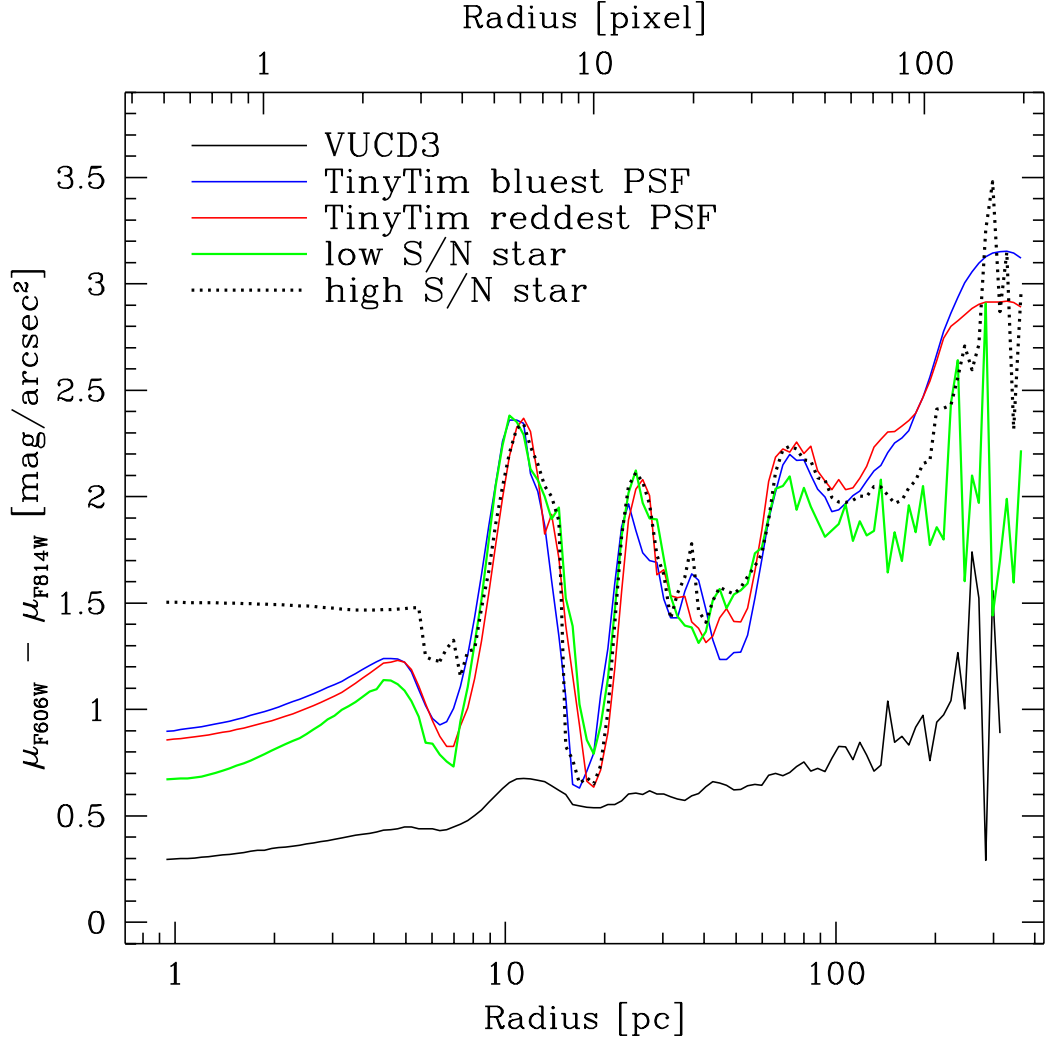}
\caption{PSF color profiles in comparison with the VUCD3 color profile (contains all PSF effects). 
All the PSF profiles are normalized at $\sim 6$ pixels ($\sim 11$ pc at the Virgo Cluster distance).}
\end{figure}
    
\begin{figure}
\epsscale{1.05}
\plotone{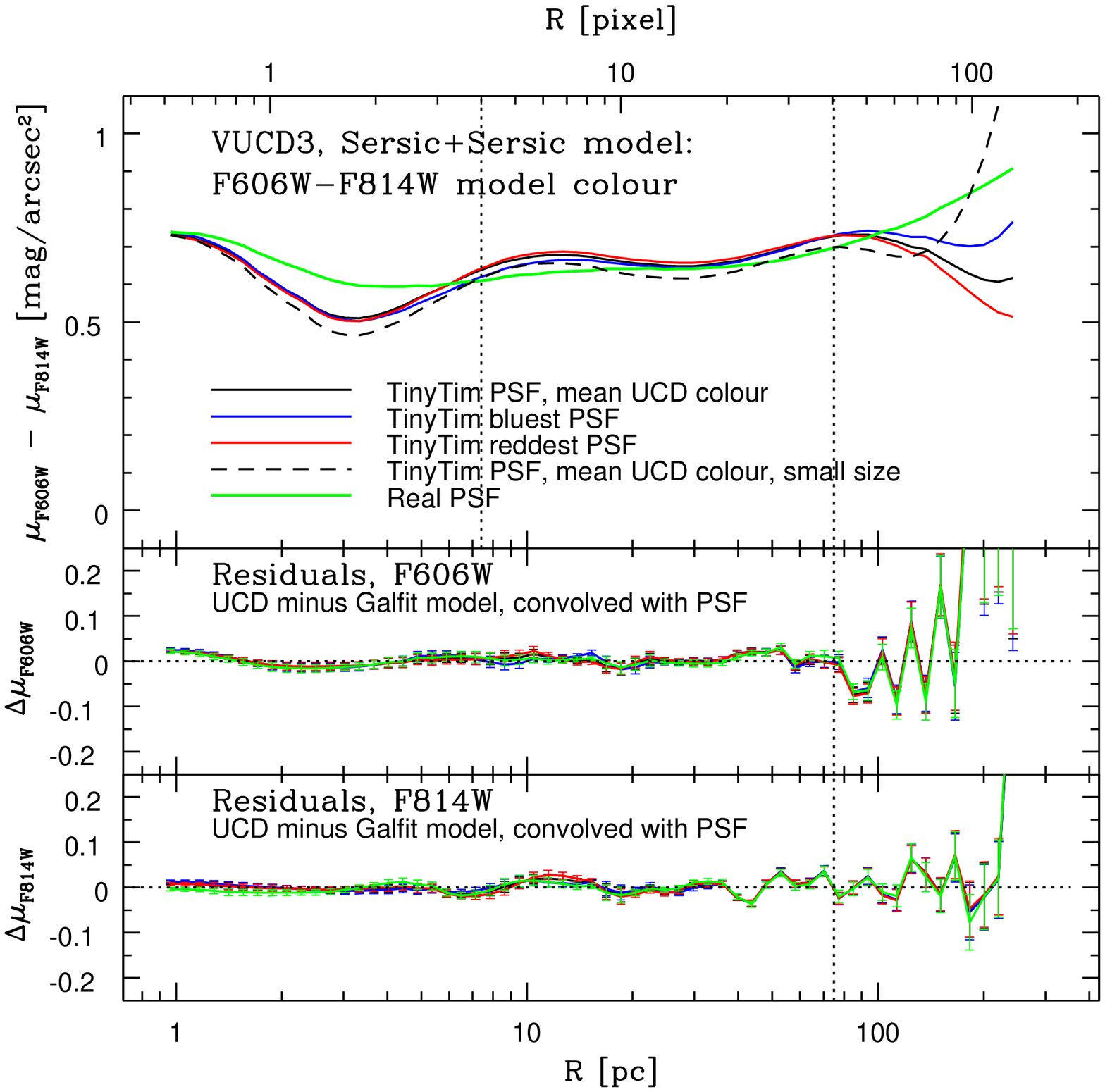}
\caption{VUCD3 color profile.}
\end{figure}

\begin{figure}
\epsscale{1.05}
\plotone{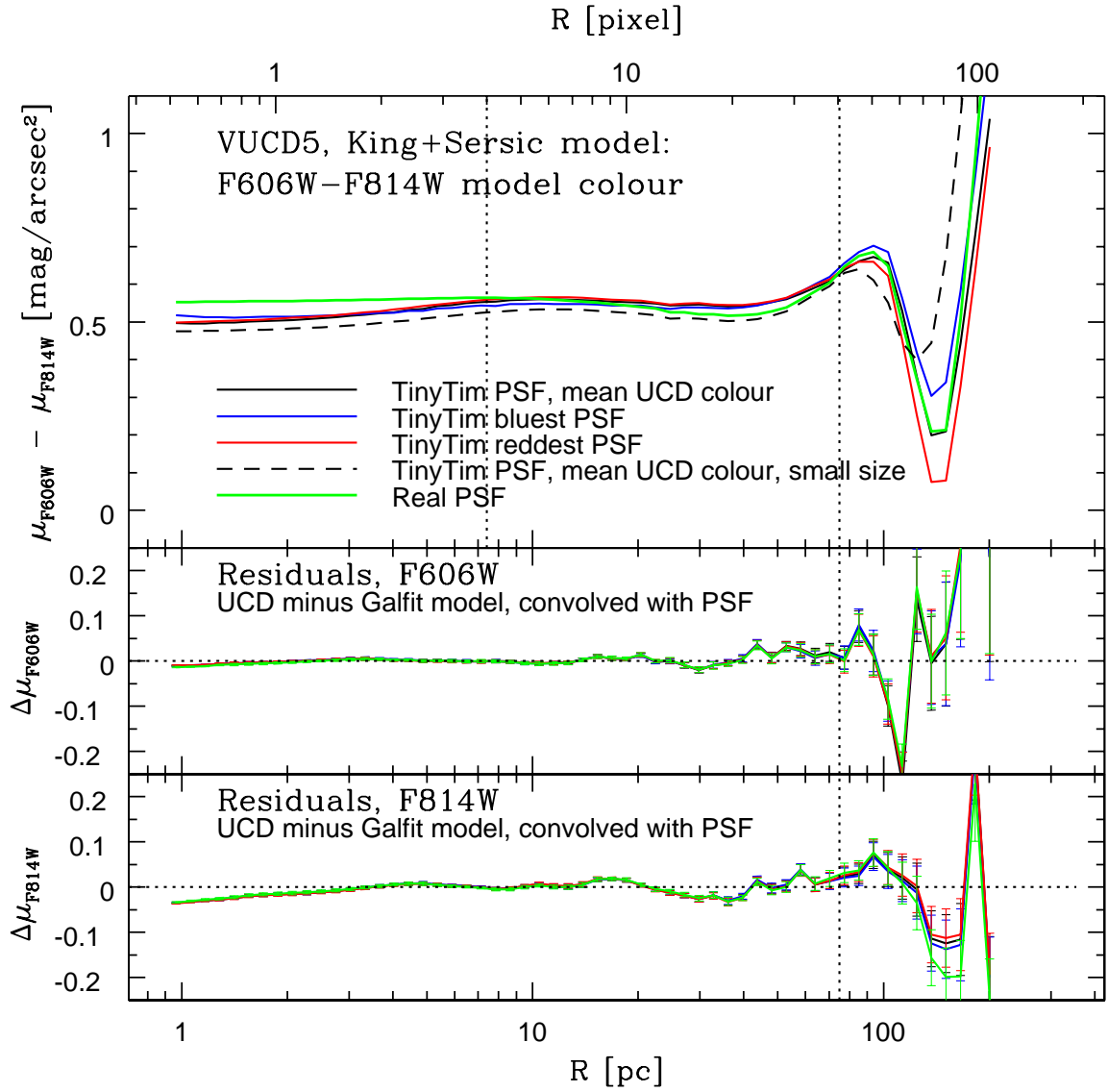}
\caption{VUCD5 color profile.}
\end{figure}

\begin{figure}
\epsscale{1.05}
\plotone{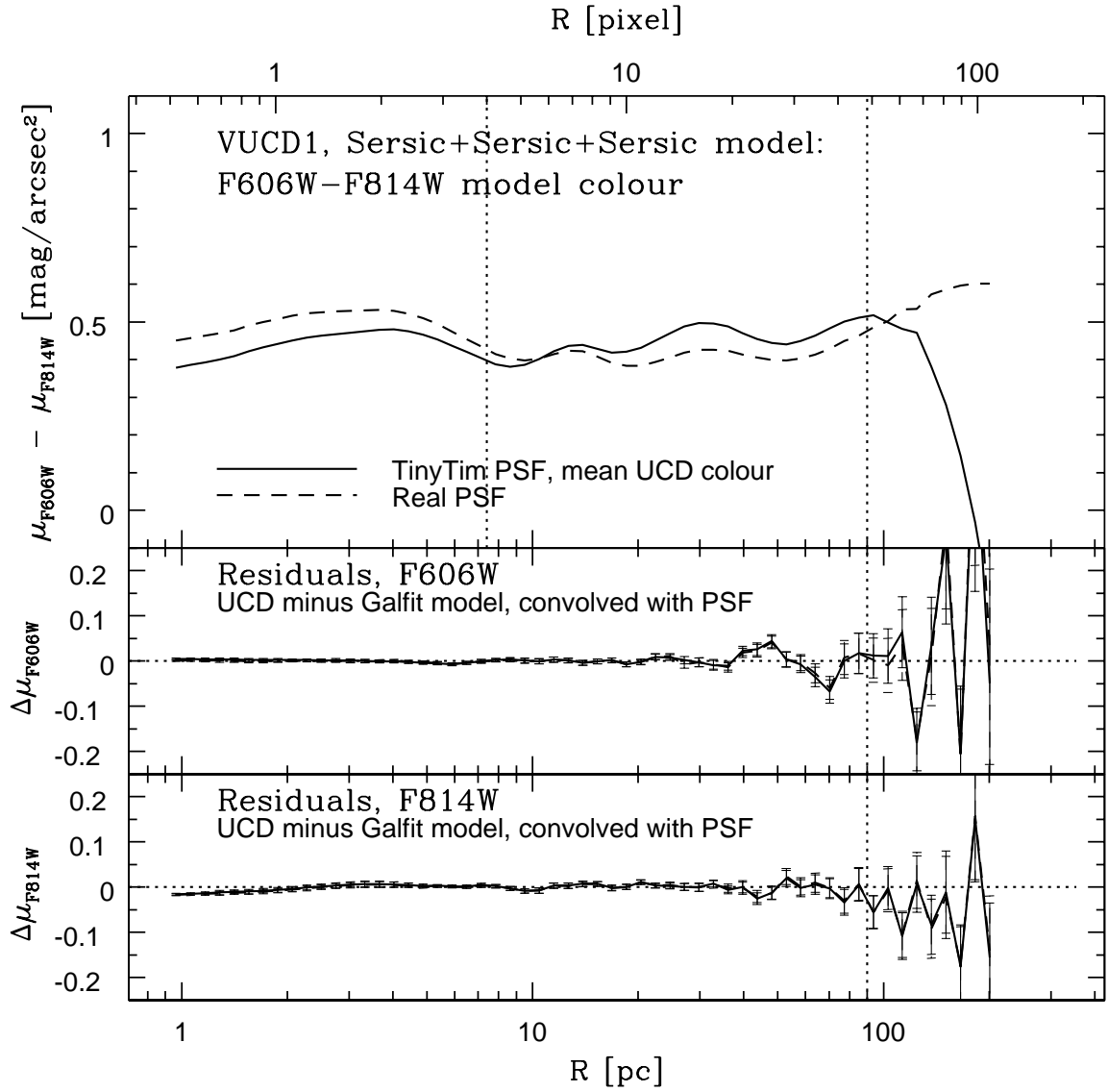}
\caption{VUCD1 color profile.}
\end{figure}

\begin{figure}
\epsscale{1.05}
\plotone{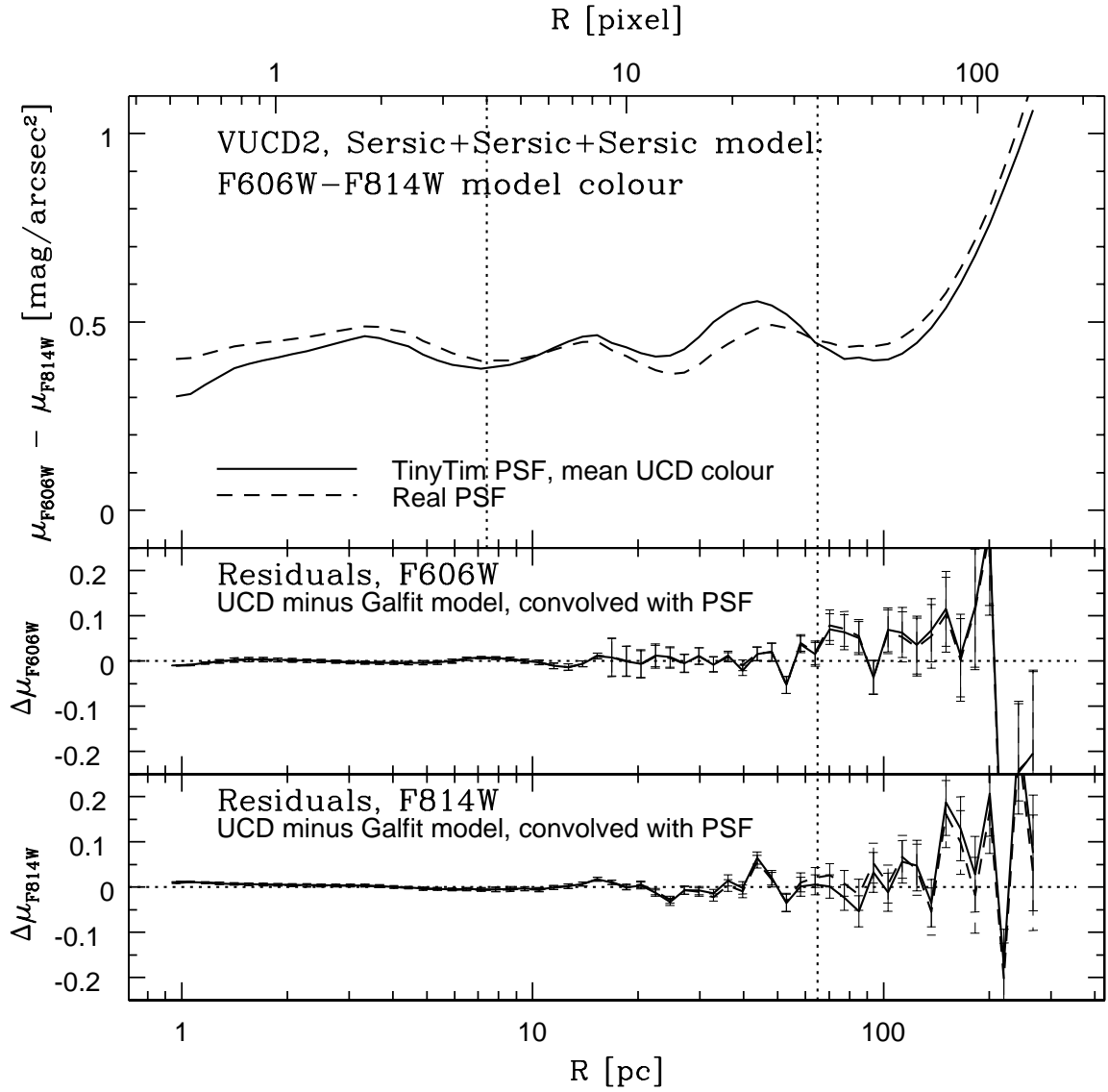}
\caption{VUCD2 color profile.}
\end{figure}

\begin{figure}
\epsscale{1.05}
\plotone{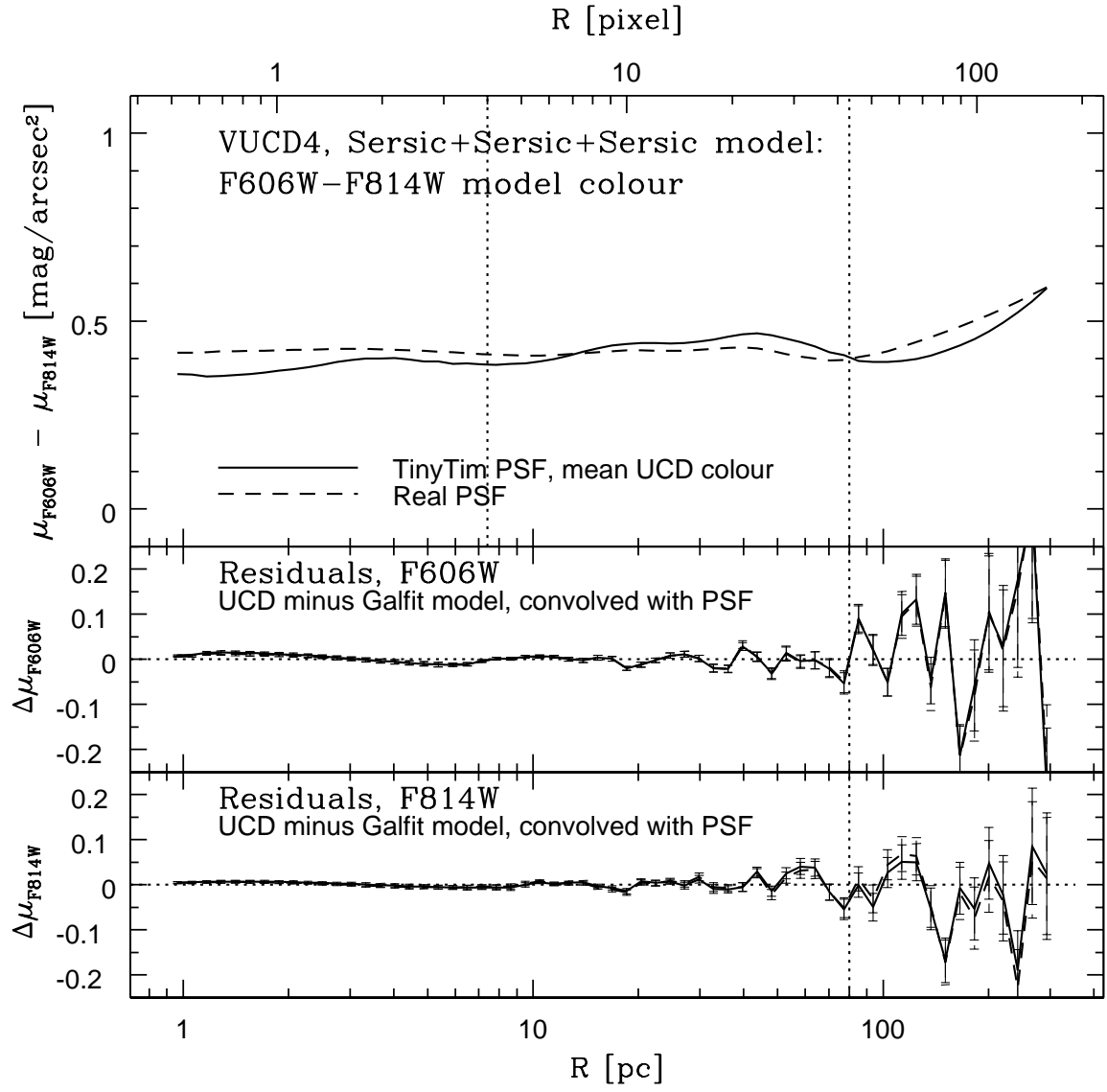}
\caption{VUCD4 color profile.}
\end{figure}

\begin{figure}
\epsscale{1.05}
\plotone{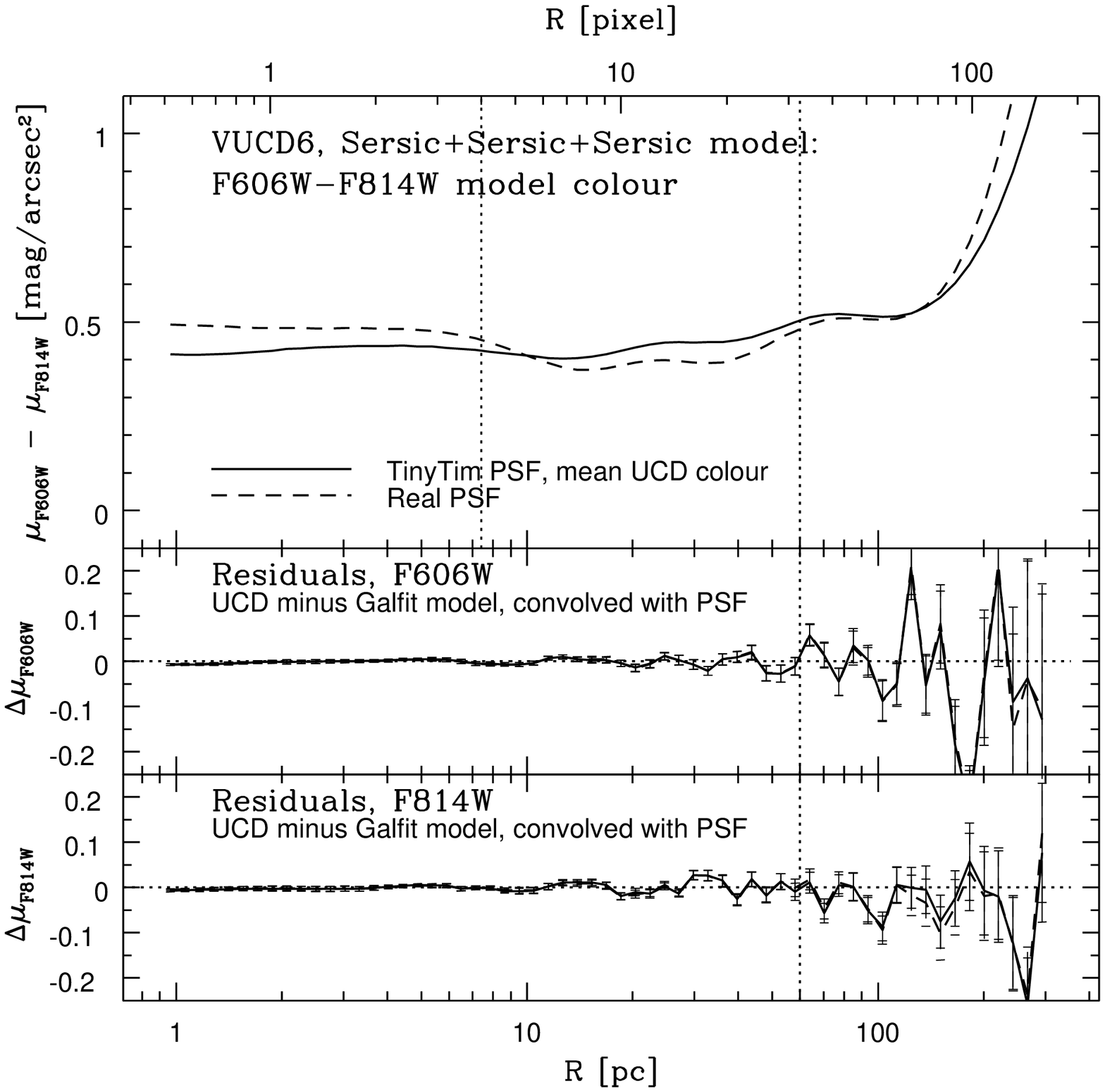}
\caption{VUCD6 color profile.}
\end{figure}

\begin{figure}
\epsscale{1.05}
\plotone{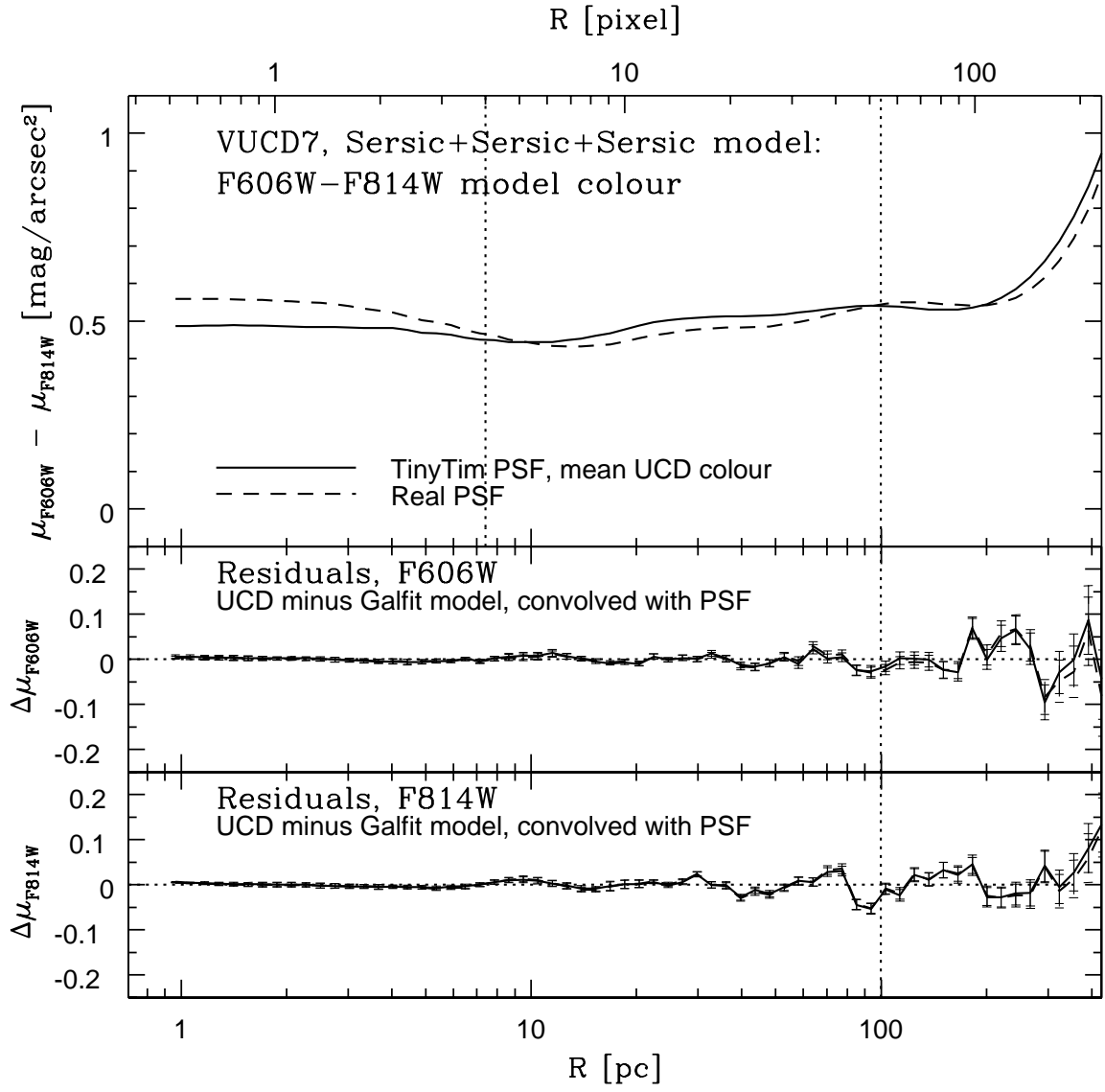}
\caption{VUCD7 color profile.}
\end{figure}

\begin{figure}
\epsscale{1.05}
\plotone{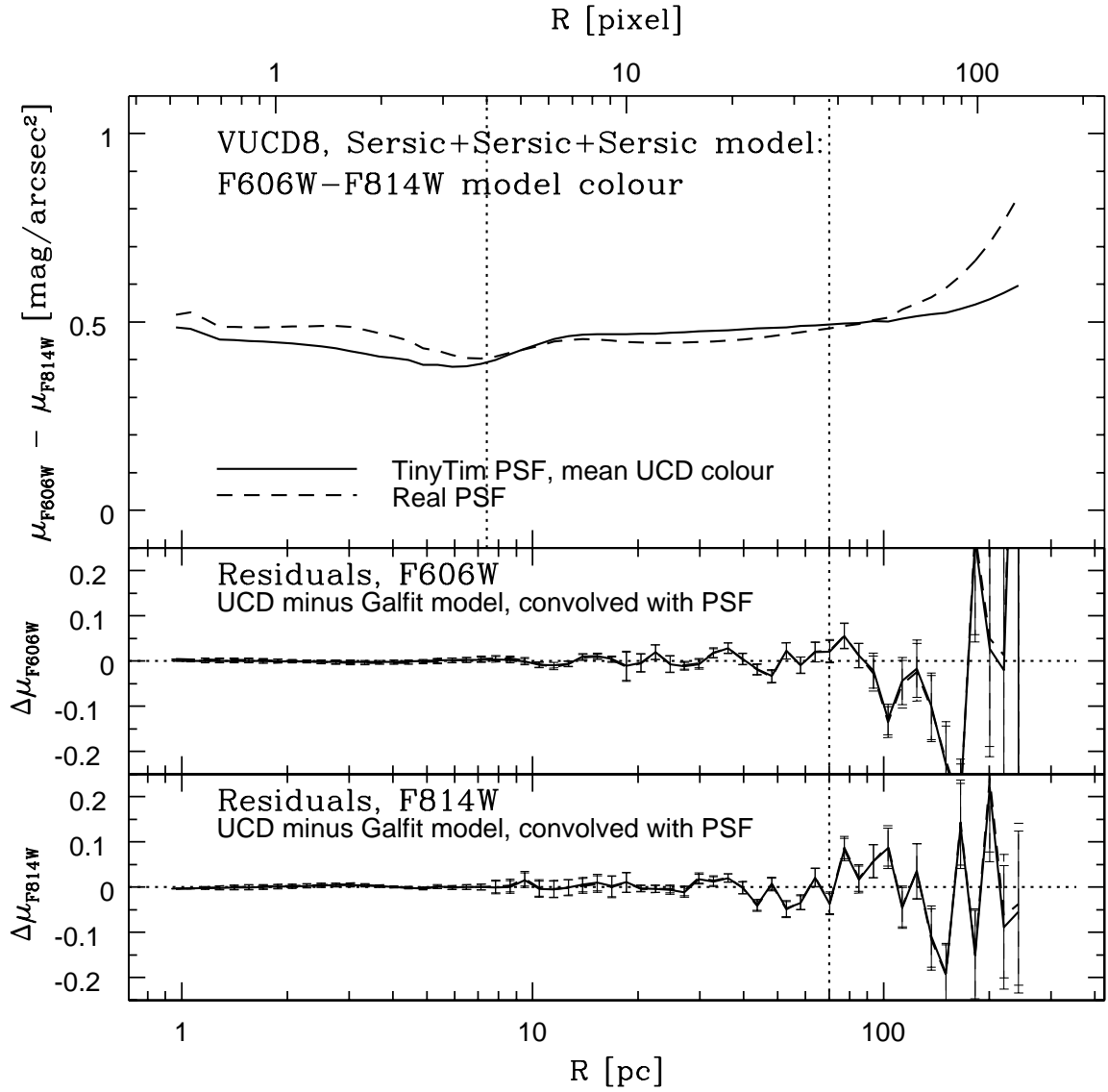}
\caption{VUCD8 color profile.}
\end{figure}

\begin{figure}
\epsscale{1.05}
\plotone{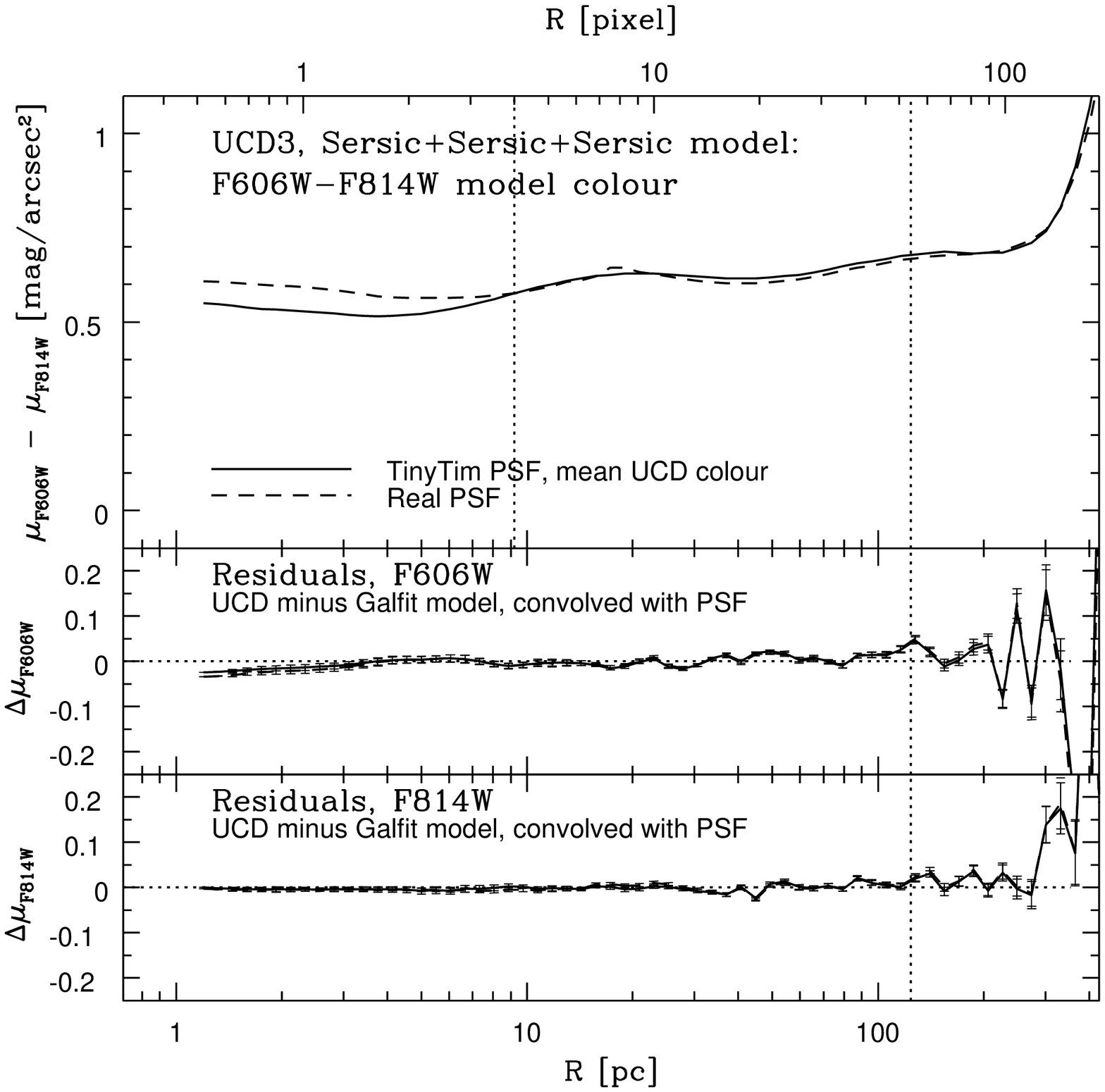}
\caption{UCD3 color profile.}
\end{figure}

\begin{figure}
\epsscale{1.05}
\plotone{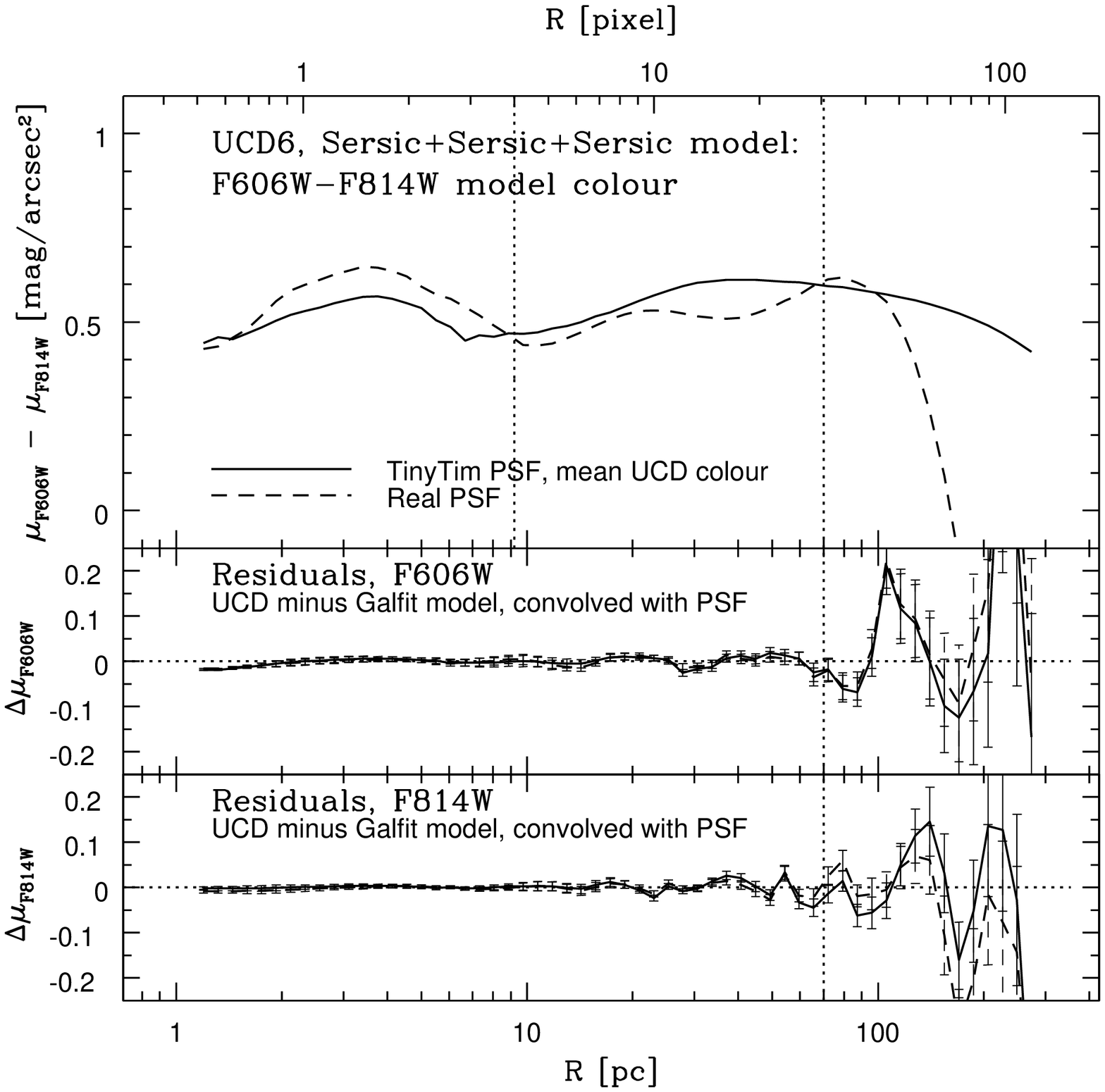}
\caption{UCD6 color profile.}
\end{figure}






\clearpage

\begin{deluxetable}{lcccccccc}
\tabletypesize{\scriptsize}
\tablecaption{UCD photometry.}
\tablewidth{0pt}
\tablehead{
\colhead{Object} & 
\colhead{R.A.(J2000)} & 
\colhead{Dec.(J2000)} & 
\colhead{$m_{V,0}$} & 
\colhead{$M_{V,0}$} & 
\colhead{$(V - I)_0$} & 
\colhead{$R_{\rm eff}$} & 
\colhead{$M_{V,0}^{\rm mod}$} & 
\colhead{$\epsilon$} \\
\colhead{} & 
\colhead{(h$\,$:$\,$m$\,$:$\,$s)} & 
\colhead{($^\circ\,$:$\,'\,$:$\,''$)} & 
\colhead{(mag)} & 
\colhead{(mag)} & 
\colhead{(mag)} & 
\colhead{(pc)} & 
\colhead{(mag)} & 
\colhead{} 
}
\startdata
UCD3 & 3:38:54.10 & -35:33:33.6 & 18.06 & -13.33 & 1.25 & 86.5 $\pm$ 6.2 & -13.55 & 0.07/0.07\,\tablenotemark{a} \\
UCD6 & 3:38:05.09 & -35:24:09.6 & 18.81 & -12.58 & 1.07 & 10.3 $\pm$ 0.9 & -12.54 & 0.12 \\
UCD16 & 3:36:47.74 & -35:48:34.1 & 20.31 & -11.08 & 1.02 & $\,$~6.4 $\pm$ 0.5 & -11.21 & 0.28 \\
UCD17 & 3:36:51.68 & -35:30:38.9 & 20.53 & -10.86 & 0.89 & 11.8 $\pm$ 0.1 & -10.91 & 0.15 \\
UCD21 & 3:37:38.29 & -35:20:20.6 & 20.90 & -10.49 & 0.94 & $\,$~7.0 $\pm$ 0.1 & -10.51 & 0.11 \\
UCD33 & 3:38:17.61 & -35:33:02.8 & 20.31 & -11.08 & 0.88 & 11.4 $\pm$ 0.0 & -11.04 & 0.08 \\
UCD41 & 3:38:29.04 & -35:22:56.5 & 19.89 & -11.50 & 1.12 & $\,$~6.9 $\pm$ 0.3 & -11.58 & 0.05 \\
UCD43 & 3:38:39.34 & -35:27:05.8 & 20.63 & -10.76 & 1.06 & $\,$~5.8 $\pm$ 0.1 & -10.67 & 0.18 \\
UCD48 & 3:39:17.72 & -35:25:30.2 & 20.11 & -11.28 & 1.19 & $\,$~5.1 $\pm$ 0.5 & -11.51 & 0.17 \\
UCD50 & 3:39:34.78 & -35:53:44.2 & 20.15 & -11.24 & 0.93 & 10.9 $\pm$ 0.7 & -11.30 & 0.32 \\
UCD52 & 3:40:19.94 & -35:15:29.8 & 20.52 & -10.87 & 1.07 & $\,$~8.7 $\pm$ 0.4 & -10.99 & 0.19 \\
UCD54 & 3:40:37.11 & -34:58:40.0 & 20.58 & -10.81 & 0.93 & $\,$~4.0 $\pm$ 0.3 & -10.62 & 0.16 \\
UCD55 & 3:41:35.88 & -35:54:57.8 & 20.30 & -11.09 & 0.97 & $\,$~9.3 $\pm$ 0.3 & -11.12 & 0.05 \\
VUCD1 & 12:30:07.61 & +12:36:31.1 & 18.68 & -12.24 & 1.01 & 11.2 $\pm$ 0.2 & -12.21 & 0.07 \\
VUCD2 & 12:30:48.24 & +12:35:11.1 & 18.69 & -12.23 & 1.05 & 13.1 $\pm$ 0.8 & -12.23 & 0.18/0.08\,\tablenotemark{a} \\
VUCD3 & 12:30:57.40 & +12:25:44.8 & 18.33 & -12.59 & 1.27 & 20.0 $\pm$ 1.5 & -12.65 & 0.16 \\
VUCD4 & 12:31:04.51 & +11:56:36.8 & 18.67 & -12.25 & 0.99 & 23.2 $\pm$ 1.4 & -12.30 & 0.16 \\
VUCD5 & 12:31:11.90 & +12:41:01.2 & 18.60 & -12.32 & 1.13 & 17.8 $\pm$ 0.3 & -12.32 & 0.01 \\
VUCD6 & 12:31:28.41 & +12:25:03.3 & 18.85 & -12.07 & 1.07 & 17.4 $\pm$ 2.4 & -12.11 & 0.04 \\
VUCD7 & 12:31:52.93 & +12:15:59.5 & 17.50 & -13.42 & 1.17 & $\,$~93.2 $\pm$ 13.1 &  -13.43 & 0.12/0.05\,\tablenotemark{a} \\
VUCD8 & 12:32:14.61 & +12:03:05.4 & 18.97 & -11.95 & 1.06 & 23.5 $\pm$ 2.5 & -11.96 & 0.16 \\ 
\multicolumn{9}{c}{Reanalysed HST/STIS data for bright Fornax UCDs from Evstigneeva et al.\ (2007):} \\
UCD1 & 3:37:03.30 & -35:38:04.6 & 19.20 & -12.19 & 1.17\,\tablenotemark{c} & 22.4 & -12.17 & 0.19 \\
UCD2 & 3:38:06.33 & -35:28:58.8 & 19.12 & -12.27 & 1.10\,\tablenotemark{c} & 23.1 & -12.28 & 0.01 \\
UCD4 & 3:39:35.95 & -35:28:24.5 & 18.94 & -12.45 & 1.07\,\tablenotemark{c} & 29.5 & -12.51 & 0.05 \\
UCD5\,\tablenotemark{b} & 3:39:52.58 & -35:04:24.1 & 19.40 & -11.99 & 0.99\,\tablenotemark{c} & 25.0 & -12.02 & 0.24/0.16\,\tablenotemark{a} \\
\enddata
\tablenotetext{a}{The first number is for the core, the second number is for the halo.}
\tablenotetext{b}{The difference to the analysis in Evstigneeva et
  al.\ (2007) is that we now derive $R_{\rm eff}$ 
and $M_{V,0}^{\rm mod}$ from a generalized King model. It gives the more stable estimate for $R_{\rm eff}$, 
than the two-component King+S\'{e}rsic model obtained in Evstigneeva et al.\ (2007).}
\tablenotetext{c}{The colors are from Karick et al.\ (2008).}
\tablecomments{The V band apparent magnitude, $m_{V,0}$, is determined as 
described in Section~3 and is corrected for foreground dust extinction 
(Schlegel et al.\ 1998). The absolute magnitude, $M_{V,0}$, is computed assuming distance 
moduli of 31.39 and 30.92~mag for the Fornax and Virgo Clusters, respectively 
(Freedman et al.\ 2001). The $(V - I)_0$ color is reddening-corrected. The half-light radius  
value, $R_{\rm eff}$, is the mean of the two pass bands, $V$ and $I$. The $R_{\rm eff}$ and $M_{V,0}^{\rm mod}$ 
values were obtained from generalized King (or standard King) models for one-component UCDs and 
King+S\'{e}rsic models for two-component UCDs (see Section~3). The ellipticity value, $\epsilon$, is 
the best model value (see last column of Table~2), mean of the two pass bands.}
\end{deluxetable}

\clearpage

\begin{deluxetable}{llcccccccl}
\tabletypesize{\scriptsize}
\tablecaption{UCD structural parameters.}
\tablewidth{0pt}
\tablehead{
\colhead{Object} & 
\colhead{S\'{e}rsic} &
\multicolumn{4}{c}{Generalized King\,\tablenotemark{a}} & 
\multicolumn{3}{c}{Nuker\,\tablenotemark{b}} & 
\colhead{Best model\,\tablenotemark{h}} \\ 
\colhead{} & 
\colhead{$n$} & 
\colhead{$\mu_{V,0}\,$\tablenotemark{c}} & 
\colhead{$R_c\,$\tablenotemark{d}} & 
\colhead{$c\,$\tablenotemark{e}} & 
\colhead{$\alpha$} & 
\colhead{$R_b\,$\tablenotemark{d}} & 
\colhead{$\beta$} & 
\colhead{$\gamma$} & 
\colhead{}
}
\startdata
UCD3\,\tablenotemark{f} & 2.12 $\pm$ 0.36 & 15.43 & $\,$~3.58 $\pm$ 0.16 & 1.68 $\pm$ 0.04 & 2.00 $\pm$ 0.00 & ... & ... & ... & K+S,S+S \\ 
UCD6 & 3.70 $\pm$ 0.18 & 13.62 & 2.57\,\tablenotemark{g} $\pm$ 0.25 & 2.31 $\pm$ 0.11 & 3.34 $\pm$ 0.18 & $\,$~3.94 $\pm$ 0.10 & 2.77 $\pm$ 0.06 & 0.17 $\pm$ 0.04 & N,GK \\ 
UCD16 & 5.17 $\pm$ 0.08 & 13.22 & 1.06\,\tablenotemark{g} $\pm$ 0.04 & 2.54 $\pm$ 0.16 & 3.00 $\pm$ 0.50 & ... & ... & ... & GK,N,S,K \\  
UCD17 & 1.08 $\pm$ 0.04  & 16.73 & $\,$~7.72 $\pm$ 0.21 & 0.91 $\pm$ 0.03 & 2.00 $\pm$ 0.00 & 37.41 $\pm$ 1.65 & 8.86 $\pm$ 0.01 & 0.01 $\pm$ 0.01 & N,S,K \\ 
UCD21 & 1.35 $\pm$ 0.01 & 15.95 & $\,$~3.77 $\pm$ 0.01 & 1.07 $\pm$ 0.01 & 2.00 $\pm$ 0.00 & 20.94 $\pm$ 0.53 & 7.40 $\pm$ 0.01 & 0.21 $\pm$ 0.02 & N,S,K,GK \\ 
UCD33 & 1.30 $\pm$ 0.04 & 16.30 & $\,$~6.89 $\pm$ 0.16 & 1.16 $\pm$ 0.02 & 2.77 $\pm$ 0.01 & 12.57 $\pm$ 0.68 & 3.80 $\pm$ 0.13 & 0.32 $\pm$ 0.01 & N,GK,S,K \\ 
UCD41 & 3.85 $\pm$ 0.01 & 13.72 & 1.30\,\tablenotemark{g} $\pm$ 0.06 & 1.78 $\pm$ 0.01 & 1.30 $\pm$ 0.05 & ... & ... & ... & GK,S,K \\ 
UCD43\,\tablenotemark{f} & 5.92 $\pm$ 1.67 & 13.51 & 0.56\,\tablenotemark{g} $\pm$ 0.06 & 1.85 $\pm$ 0.12 & 2.00 $\pm$ 0.00 & 10.68 $\pm$ 1.00 & 2.76 $\pm$ 0.28 & 1.68 $\pm$ 0.06 & K+S,N,S,GK \\ 
UCD48 & 8.24 $\pm$ 0.43 & 11.58 & 0.47\,\tablenotemark{g} $\pm$ 0.07 & 2.33 $\pm$ 0.08 & 1.10 $\pm$ 0.10 & 34.37 $\pm$ 1.80 & 3.65 $\pm$ 0.08 & 1.76 $\pm$ 0.02 & N,GK,S,K \\ 
UCD50 & 3.25 $\pm$ 0.07 & 14.79 & $\,$~2.84 $\pm$ 0.06 & 2.29 $\pm$ 0.06 & 3.56 $\pm$ 0.04 & $\,$~8.43 $\pm$ 0.01 & 3.68 $\pm$ 0.05 & 0.00 $\pm$ 0.00 & N,K,GK \\ 
UCD52 & 4.21 $\pm$ 0.02 & 14.52 & 1.56\,\tablenotemark{g} $\pm$ 0.06 & 1.82 $\pm$ 0.08 & 1.28 $\pm$ 0.23 & 15.39 $\pm$ 1.15 & 2.87 $\pm$ 0.07 & 1.42 $\pm$ 0.06 & N,S \\ 
UCD54 & 3.47 $\pm$ 0.26 & 13.87 & 0.78\,\tablenotemark{g} $\pm$ 0.19 & 1.73 $\pm$ 0.01 & 1.14 $\pm$ 0.41 & ... & ... & ... &  N,S,K,GK \\ 
UCD55 & 2.15 $\pm$ 0.06 & 15.57 & $\,$~3.49 $\pm$ 0.09 & 1.57 $\pm$ 0.03 & 2.55 $\pm$ 0.28 & $\,$~8.90 $\pm$ 0.57 & 3.88 $\pm$ 0.04 & 0.01 $\pm$ 0.00 & K,GK,S,K \\ 
VUCD1 & 2.23 $\pm$ 0.11 & 14.67 & $\,$~4.28 $\pm$ 0.11 & 1.92 $\pm$ 0.06 & 3.74 $\pm$ 0.03 & $\,$~6.25 $\pm$ 0.14 & 2.69 $\pm$ 0.03 & 0.23 $\pm$ 0.00 & N \\ 
VUCD2\,\tablenotemark{f} & 1.11 $\pm$ 0.10 & 15.09 & $\,$~5.16 $\pm$ 0.12 & 0.94 $\pm$ 0.04 & 2.00 $\pm$ 0.00 & $\,$~7.56 $\pm$ 0.07 & 2.61 $\pm$ 0.06 & 0.54 $\pm$ 0.01 & K+S,S+S \\ 
VUCD3 & 8.27 $\pm$ 0.06 & 13.80 & $\,$~2.10 $\pm$ 0.15 & 2.06 $\pm$ 0.00 & 0.62 $\pm$ 0.01 & ... & ... & ... &  S,N \\ 
VUCD4\,\tablenotemark{f} & 1.11 $\pm$ 0.11 & 16.01 & $\,$~7.52 $\pm$ 0.09 & 0.96 $\pm$ 0.05 & 2.00 $\pm$ 0.00 & $\,$~9.08 $\pm$ 0.16 & 2.35 $\pm$ 0.04 & 0.32 $\pm$ 0.04 & N,K+S,S+S \\ 
VUCD5 & 1.91 $\pm$ 0.04 & 15.97 & $\,$~6.56 $\pm$ 0.01 & 1.42 $\pm$ 0.01 & 2.00 $\pm$ 0.00 & 18.21 $\pm$ 1.39 & 3.92 $\pm$ 0.16 & 0.00 $\pm$ 0.00 & K,N \\ 
VUCD6\,\tablenotemark{f} & 1.16 $\pm$ 0.11 & 15.26 & $\,$~4.08 $\pm$ 0.06 & 0.96 $\pm$ 0.02 & 2.00 $\pm$ 0.00 & $\,$~5.21 $\pm$ 0.35 & 2.24 $\pm$ 0.04 & 0.50 $\pm$ 0.06 & N,K+S,S+S \\ 
VUCD7\,\tablenotemark{f} & 2.18 $\pm$ 0.10 & 14.38 & $\,$~3.09 $\pm$ 0.04 & 1.62 $\pm$ 0.04 & 2.00 $\pm$ 0.00 & ... & ... & ... &  K+S,S+S \\ 
VUCD8 & 4.07 $\pm$ 0.22 & 15.41 & $\,$~3.33 $\pm$ 0.04 & 2.01 $\pm$ 0.08 & 1.28 $\pm$ 0.00 & $\,$~3.69 $\pm$ 0.11 & 2.24 $\pm$ 0.04 & 0.00 $\pm$ 0.00 & N,K,GK 
\enddata
\tablenotetext{a}{If the standard King model ($\alpha = 2.0$) fits the object
better than the generalized King model ($\alpha$ is a free parameter), we 
provide parameters for the standard King model.}
\tablenotetext{b}{The Nuker model parameters are provided only if this model is reasonably good for the 
object and there is a good agreement between two pass bands.}
\tablenotetext{c}{$\mu_{V,0}$ is measured in mag arcsec$^{-2}$.}
\tablenotetext{d}{$R_c$ and $R_b$ are measured in pc.}
\tablenotetext{e}{Concentration $c={\rm log}\,(R_t/R_c)$.}
\tablenotetext{f}{For two-component UCDs, S\'{e}rsic and King model parameters are given for the central (core) component.}
\tablenotetext{g}{$R_c < 1$ pix, core is unresolved.}
\tablenotetext{h}{N -- Nuker, K -- King ($\alpha=2$), GK -- Generalized King ($\alpha$ -- any), S -- S\'{e}rsic, 
K+S -- King plus S\'{e}rsic, S+S -- S\'{e}rsic plus S\'{e}rsic.}
\tablecomments{All the parameters (except $\mu_{V,0}$) are the means of the two pass bands, $V$ and $I$.}
\end{deluxetable}


\end{document}